\documentclass[transmag]{IEEEtran}
\usepackage{latexsym}
\usepackage{graphicx}
\usepackage{amsfonts,amssymb,amsmath}
\usepackage{hyperref}
\usepackage{algorithm}
\usepackage{algpseudocode}
\usepackage{textcomp}
\usepackage{xcolor}
\usepackage{hyperref}
\usepackage{ulem}
\usepackage{float}
\usepackage{caption}
\usepackage{subcaption}
\def\BibTeX{{\rm B\kern-.05em{\sc i\kern-.025em b}\kern-.08em T\kern-.1667em\lower.7ex\hbox{E}\kern-.125emX}}
\begin{document}

\title{HierTrain: Fast Hierarchical Edge AI Learning with Hybrid Parallelism in
 Mobile-Edge-Cloud Computing}

\author{Deyin Liu, Xu Chen,  Zhi Zhou, and Qing Ling
\thanks{The authors are with School of Data and Computer Science, Sun Yat-sen
 University, Guangzhou 510006, China (e-mail:\ liudy7@mail2.sysu.edu.cn;
\{chenxu35, zhouzhi9, lingqing556\}@mail.sysu.edu.cn).}}

\IEEEtitleabstractindextext{\begin{abstract}
Nowadays, deep neural networks (DNNs) are the core enablers for many emerging edge AI applications. Conventional approaches to training DNNs are generally implemented at central servers or cloud centers for centralized learning, which is typically time-consuming and resource-demanding due to the transmission of a large amount of data samples from the device to the remote cloud. To overcome these disadvantages, we consider accelerating the learning process of DNNs on the Mobile-Edge-Cloud Computing (MECC) paradigm. In this paper, we propose HierTrain, a hierarchical edge AI learning framework, which efficiently deploys the DNN training task over the hierarchical MECC architecture. We develop a novel \textit{hybrid parallelism} method, which is the key to HierTrain, to adaptively assign the DNN model layers and the data samples across the three levels of edge device, edge server and cloud center. We then formulate the problem of scheduling the DNN training tasks at both layer-granularity and sample-granularity. Solving this optimization problem enables us to achieve the minimum training time. We further implement a hardware prototype consisting of an edge device, an edge server and a cloud server, and conduct extensive experiments on it. Experimental results demonstrate that HierTrain can achieve up to 6.9$\times$ speedup compared to the cloud-based hierarchical training approach.
\end{abstract}

}

\maketitle

\section{INTRODUCTION}

In recent years, deep learning has become a popular research topic and been integrated into a large number of applications, including image recognition \cite{simonyan2014very}, natural language processing \cite{devlin2018bert}, recommendation systems \cite{covington2016deep}, to name a few. Moreover, empowered by edge computing, many real-time deep learning based edge AI applications are emerging in various domains such as smart healthcare, smart robots and industrial IoT\cite{zhou2019edge}.

As a data-driven approach, deep learning based edge AI typically requires to have adequate data samples, from which deep neural networks (DNNs) are trained to extract features or attributes. These data samples are often generated by mobile and IoT devices at the network edge that have limited communication and computation capabilities, such as mobile phones, smart watches, smart robots, etc. Therefore, how to efficiently utilize the communication and computation capabilities of edge devices to train DNNs with the generated data samples will be a vital issue for many emerging edge AI applications.

One solution to this problem is cloud computing \cite{kumar2019comprehensive,zhang2017poseidon}, which allows edge devices to offload their data samples to a cloud center. Then, the resource-intensive task of training a DNN is conducted in the cloud center, often implemented in parallel on multiple computing units. Despite cloud computing provides almost unlimited computation resources, the major concern comes from the high data transmission latency and overhead over the Internet, which slows down the training process and hinders the real-time model update. Another solution is to train the DNN in a fully decentralized peer-to-peer manner \cite{mathur2018hydra}. This approach avoids the communication overhead between the edge devices and the cloud center. Nevertheless, when the computation resources of the edge devices are limited, solely relying them to train the DNN is impractical, or may cause significant computation delay. We classify these two approaches as \textit{horizontal training}, as the computation tasks are executed over multiple workers at the same system level (either the computing units in the cloud center, or the edge devices in the fully decentralized peer-to-peer network).
\begin{figure}[t!]
  \centering
  \includegraphics[scale=0.5]{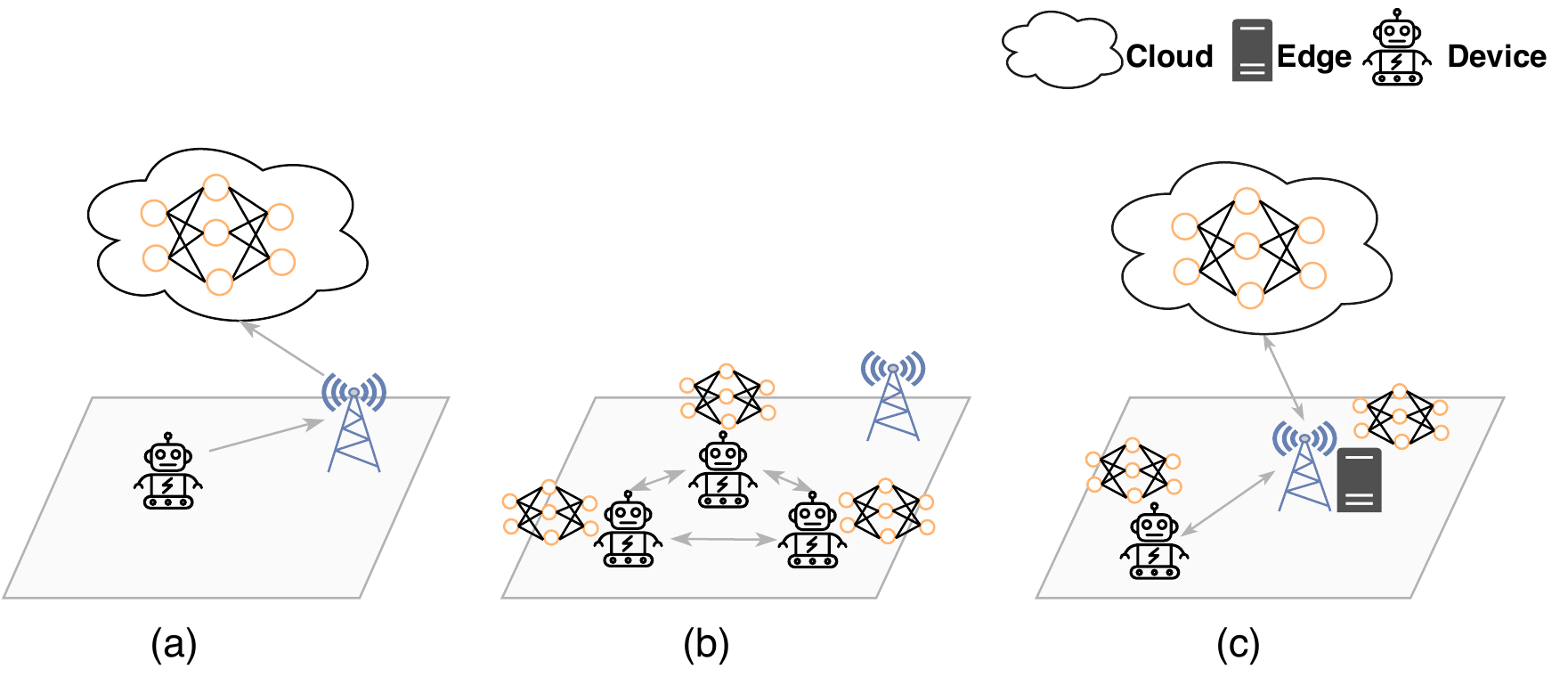}
  \caption{Training a DNN: (a) on the cloud center; (b) on the edge devices in a fully decentralized peer-to-peer manner; (c) on the mobile-edge-cloud hierarchical architecture. Here (a) and (b) belong to \textit{horizontal training}, while (c) is \textit{hierarchical training}.}
  \label{fig:high_level}
\end{figure}

\begin{figure*}[!t]
  \centering
  \begin{subfigure}[b]{0.33\textwidth}
    \centering
    \includegraphics[scale=0.38]{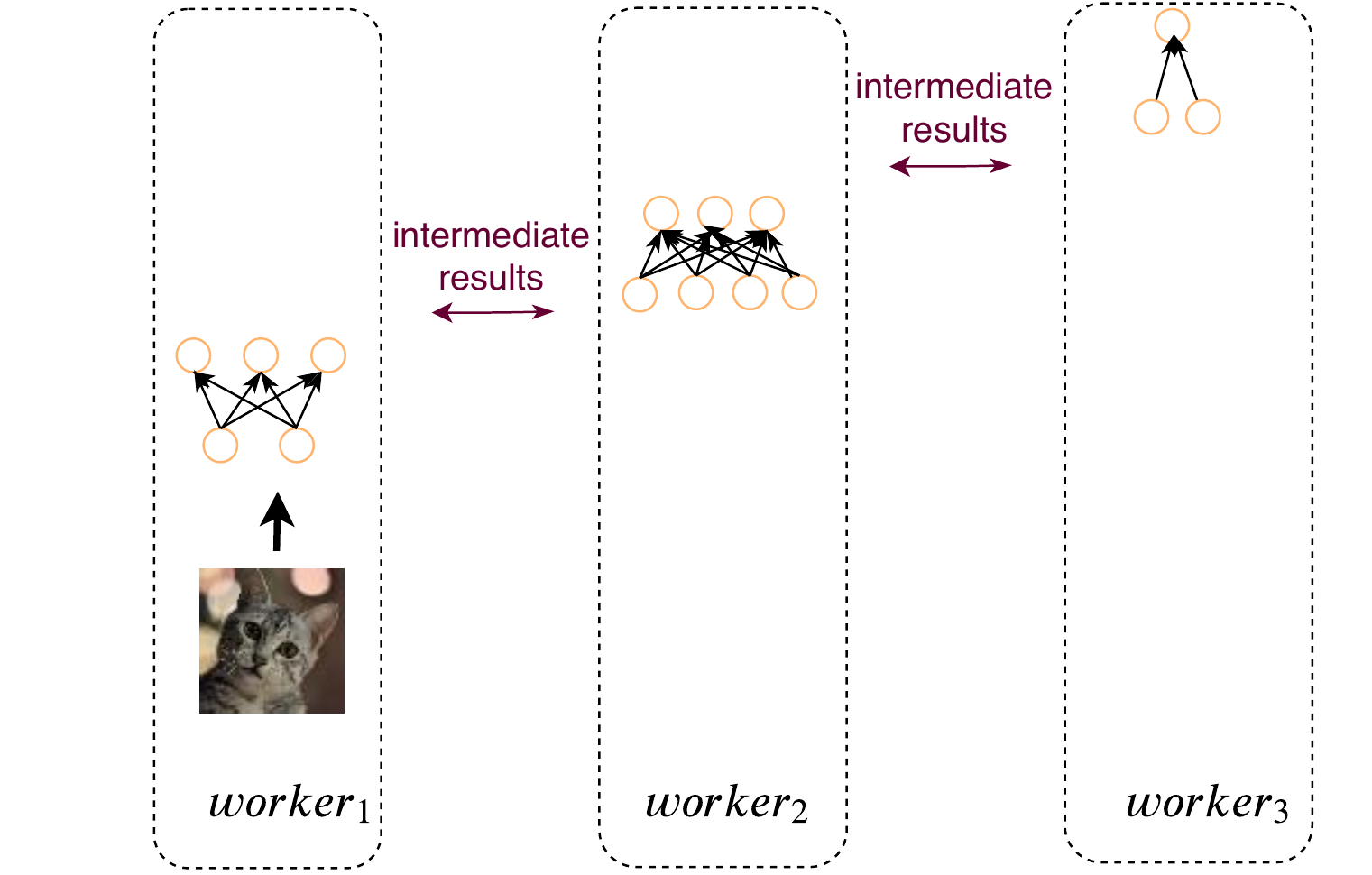}
    \caption{Model Parallelism}
    \label{fig:sequence}
    \end{subfigure}
  \begin{subfigure}[b]{0.32\textwidth}
    \centering
    \includegraphics[scale=0.38]{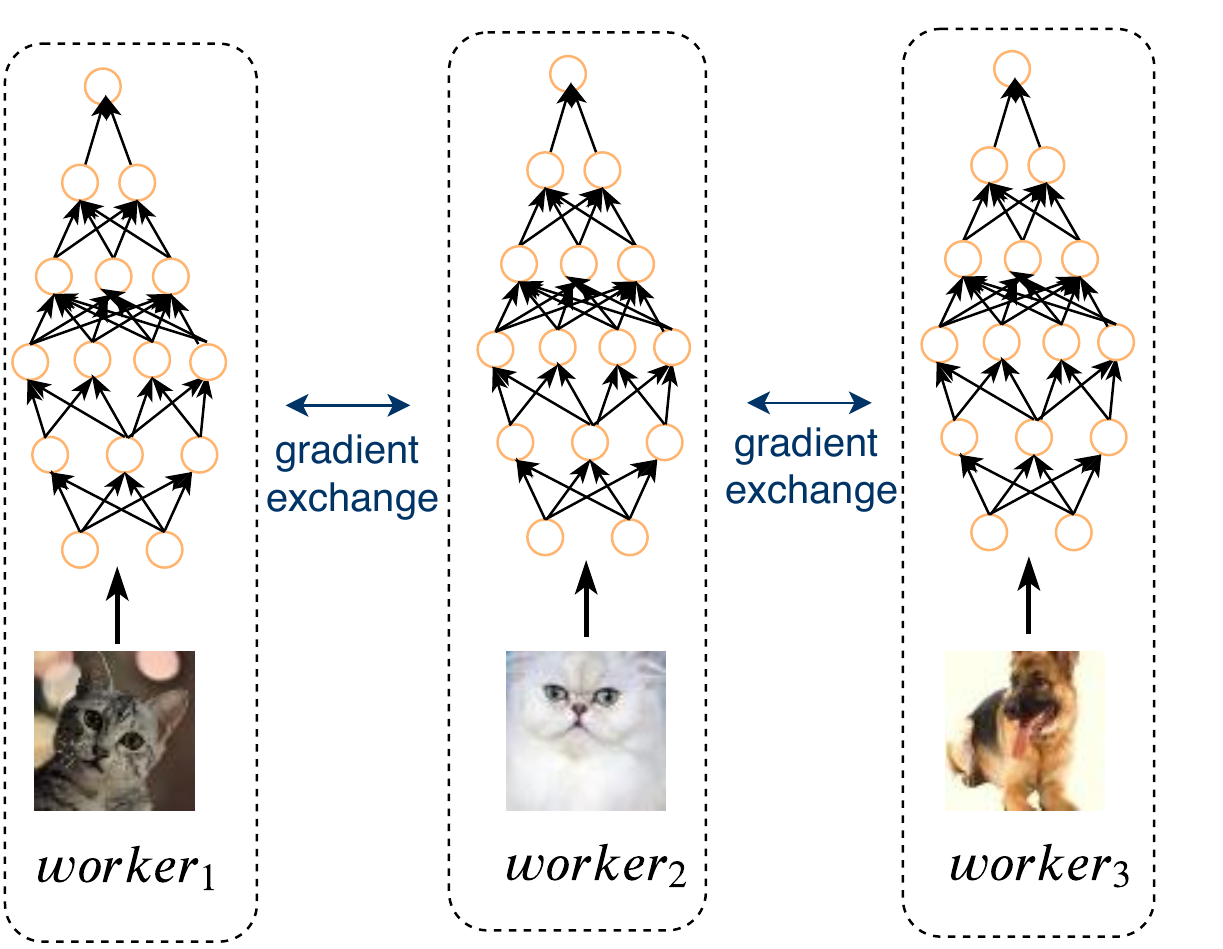}
    \caption{Data Parallelism}
    \label{fig:data}
    \end{subfigure}
  \begin{subfigure}[b]{0.33\textwidth}
    \centering
    \includegraphics[scale=0.38]{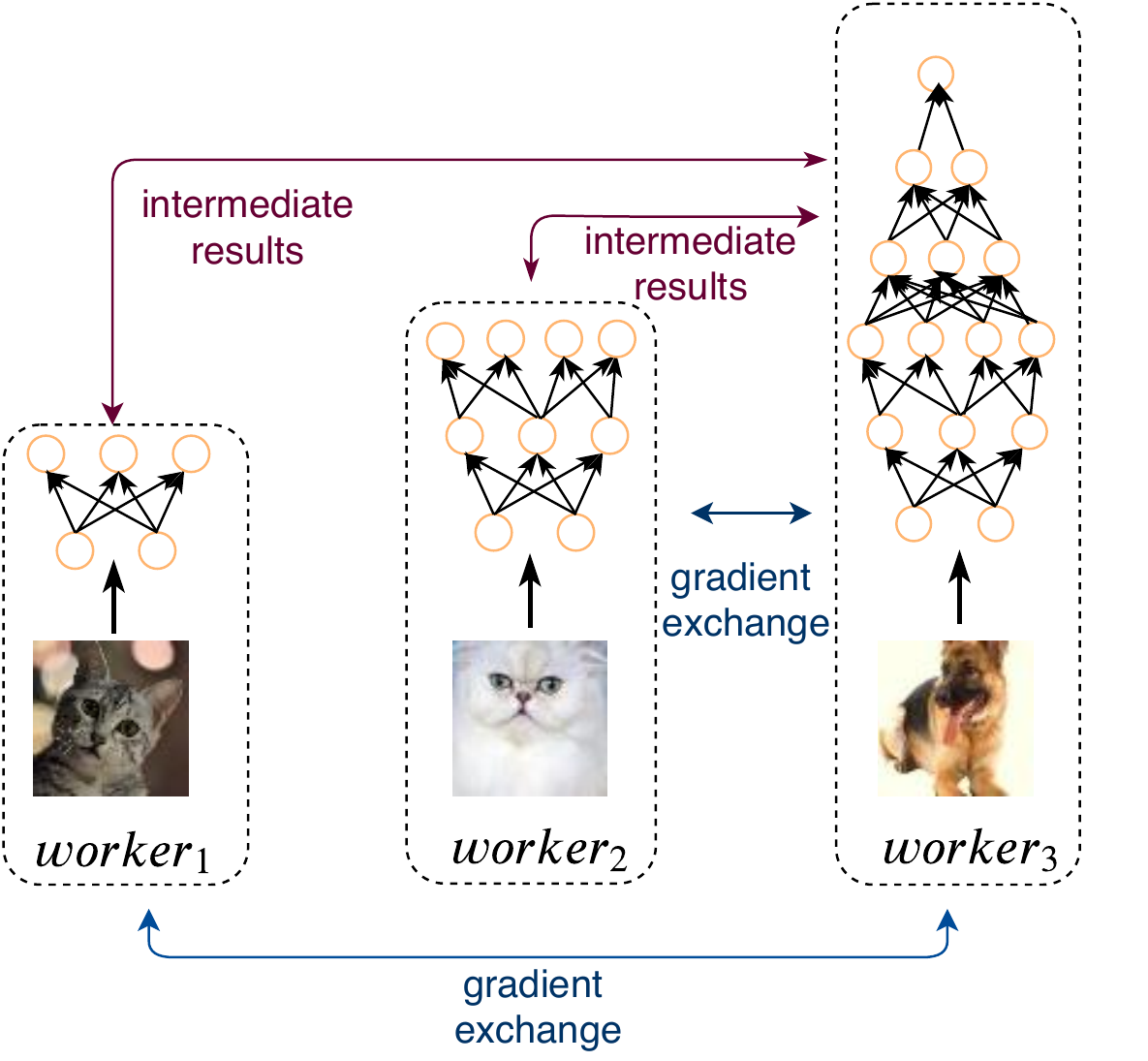}
    \caption{Hybrid Parallelism}
    \label{fig:hybrid}
    \end{subfigure}
    \caption{Illustration of the three parallelism methods. Each row of circles represents a layer in the trained DNN.}
    \label{fig:all_parallelism}
\end{figure*}

There are also \textit{hierarchical training} approaches to efficient training of DNNs. JointDNN is proposed in \cite{eshratifar2018jointdnn}, which trains some layers of a DNN on an edge device and the other layers on the cloud center. However, the latency between the edge device and the cloud center is still the major factor to limit the training speed. The emerging edge computing paradigm provides another option, in which the edge servers are in between of the edge devices and the cloud center, and can fulfill computation tasks as close as possible to the data sources. Comparing to the communication latency between the cloud center and the edge devices, that between the edge servers and the edge devices is much lower. These excellent properties motivate the emerging edge learning scheme of jointly training a DNN with an edge device and an edge server \cite{ren2019accelerating}. Fig. \ref{fig:high_level} illustrates the difference between the \textit{horizontal training} and \textit{hierarchical training} paradigms. Observing that the works in \cite{eshratifar2018jointdnn} and  \cite{ren2019accelerating} only consider two levels in the mobile-edge-cloud hierarchical architecture -- the device and cloud levels in \cite{eshratifar2018jointdnn} and the mobile and edge levels in \cite{ren2019accelerating}, the drawback of them is that they did not fully utilize the communication and computation resources of all the three levels. As communication latency between mobile and edge levels is generally low and the computation resource at the cloud level is abundant, a holistic framework that fully exploits the communication and computation resources of all the three levels can definitely leash the great potentials of mobile-edge-cloud computing for accelerating edge AI learning.

 Motivated by this, we propose a hierarchical training framework, abbreviated as HierTrain, which efficiently deploys the DNN training tasks over the mobile-edge-cloud levels and achieves minimum training time for fast edge AI learning. Our contributions are summarized as follows. 
 
 \begin{enumerate}[\IEEEsetlabelwidth{12)}]
\item We develop a novel \textit{hybrid parallelism} method, which is the key to
  HierTrain, to adaptively assign the DNN model layers and the data samples to
  the three levels by taking into account the communication and computation resource heterogeneity therein. 
\item We formulate the problem of scheduling the DNN training tasks at both layer-granularity and sample-granularity. Solving this minimization problem enables us to achieve the minimum training time.
\item We implement and deploy a hardware prototype over an edge device, an edge server and a cloud server, and extensive experimental results demonstrate that HierTrain achieves superior performance, e.g., achieving up to 6.9$\times$ speedup compared to the cloud-based hierarchical training approach.
\end{enumerate}

We should emphasize that, different from many existing works focusing on edge AI inference \cite{li2019edge}, in this study we promote HireTrain for addressing the important issue of edge AI training acceleration. This is due to the emerging demand that many edge AI applications (e.g., smart robots and industrial IoT) require both real-time performance and continuous learning capability of fast model updating with fresh sensing/input data samples and being adaptive to complex dynamic application environments. On the other hand, HierTrain is along the emerging line of promoting in-network model training such as edge learning for intelligent B5G networking \cite{murshed2019machine} for mitigating the significant overhead and latency of transferring the data of massive size to the cloud for remote model training.

\section{Background \& Motivation}

In general, there are three computing workers/nodes for DNN training in the mobile-edge-cloud hierarchical system: edge device, edge server and cloud center, which have diverse communication and computation capacities. To jointly train a DNN, we need to determine how to split the training data samples and the trained DNN across the three workers. Below, we introduce two traditional methods, \textit{model parallelism} and \textit{data parallelism}, as well as our proposed \textit{hybrid parallelism} method. The three parallelism methods are illustrated in Fig. \ref{fig:all_parallelism}.

\textit{1) Model Parallelism:} Because a DNN is typically stacked by a sequence of distinct layers, it is
natural to assign the layers to the workers; see Fig. \ref{fig:sequence}. In the
\textit{model parallelism} method, each worker holds multiple layers and is in
charge of updating the corresponding model parameters. Therefore, when training the DNN with the
back-propagation rule in the stochastic gradient descent (SGD) algorithm \cite{bottou2010large}, the workers need to communicate to exchange the intermediate results. The works of JointDNN \cite{eshratifar2018jointdnn} and JALAD \cite{li2018jalad} demonstrate the effectiveness of the \textit{model parallelism} method. However, since the layers of the DNN are trained sequentially, when one worker is computing the others must stay idle. Thus, the computation resources are not fully utilized in the \textit{model parallelism} method.

\textit{2) Data Parallelism:} The \textit{data parallelism} method splits the data samples to the workers, trains one local copy of DNN in every worker, and forces the local DNNs to reach a consensus along the optimization process. To implement SGD, the workers need to exchange either the local stochastic gradients or the local model parameters from time to time, as depicted in Fig. \ref{fig:data}. The works of \cite{goyal2017accurate} and \cite{you2017large} show that the \textit{data parallelism} method is able to accelerate the DNN training when the data are collected and split to multiple computing units within the cloud center. Nevertheless, the requirement of transmitting the local stochastic gradients or the local model parameters, whose dimensions are the same, leads to heavy communication overhead when the size of DNN is large. Therefore, the \textit{data parallelism} method is not communication-efficient in the mobile-edge-cloud architecture.

\begin{figure*}[!t]
 \centering
 \includegraphics[scale=0.48]{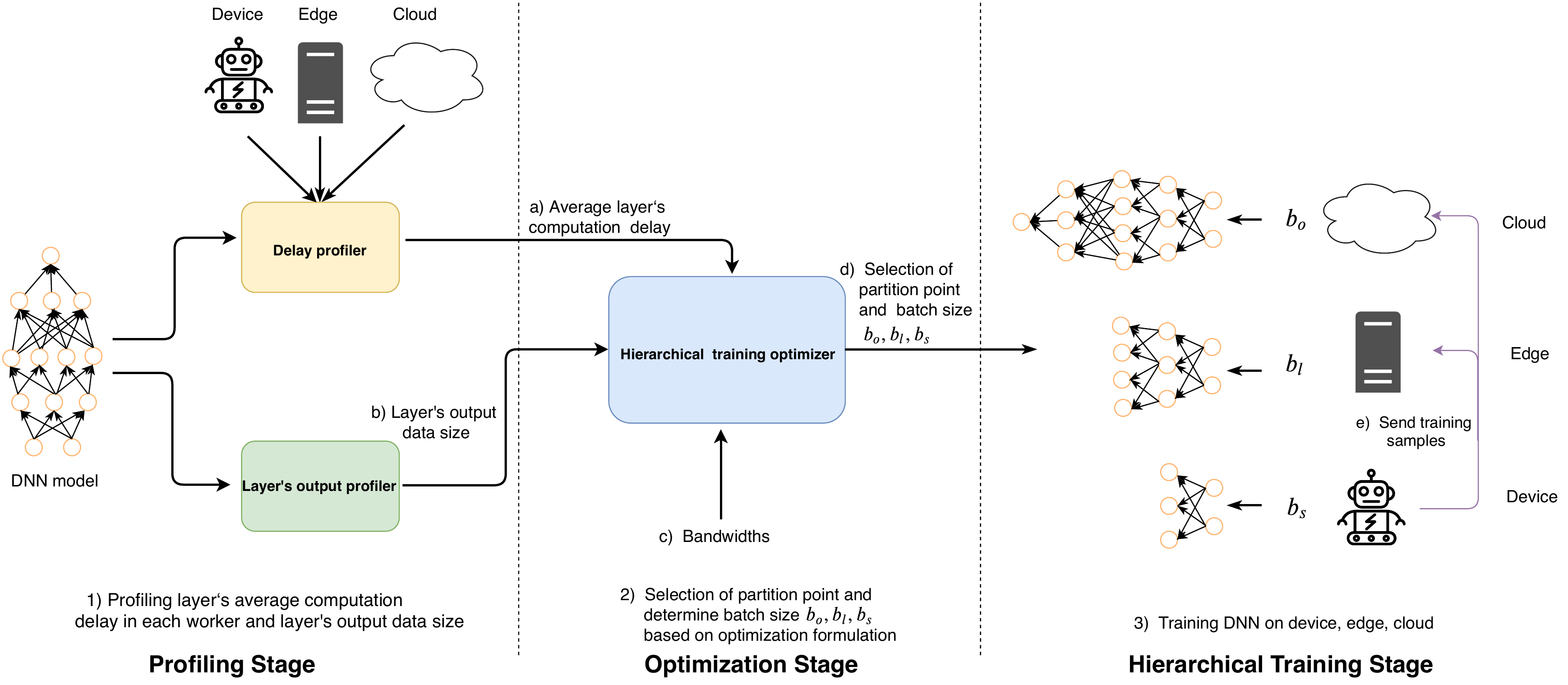}
 \caption{System overview.}
 \label{fig:systems}
\end{figure*}

\textit{3) Hybrid Parallelism:} Observe that the backend layers in most DNNs, such as convolutional neural networks (CNNs), are fully connected layers and contain the majority of
parameters. This fact motivates us to improve the \textit{model parallelism} method through letting all the backend layers be trained by one worker while the frontend layers be trained by multiple workers. Therefore, the workers just need
to exchange a small fraction of the local stochastic gradients or the local model parameters to train the frontend layers, as well as transmit the intermediate results to train the backend layers, thus the communication latency between workers is largely reduced. As shown in Fig. \ref{fig:hybrid}, the backend layers are only trained by $worker_3$. Some frontend layers are trained by $worker_2$ and $worker_3$, while some are jointly trained by all the workers. Meanwhile, similar to the data parallelism method, training data samples are split and assigned to all the workers according to their computing resource heterogeneity, to further balance the workloads across the device, edge and cloud. 

In order to apply the \textit{hybrid parallelism} method to accelerate the training of DNNs over the mobile-edge-cloud architecture, we need to optimize the assignments of the DNN layers and the data samples to the three workers. To this end, we propose HierTrain, a hierarchical training framework, as follows.

\section{HierTrain Framework}
In this section, we present the HierTrain framework, which jointly selects the best partition point of the given DNN model and determines the appropriate number of data samples delegated to different workers in a mobile-edge-cloud hierarchy. Fig. \ref{fig:systems} presents the system overview of the HierTrain framework, which consists of three stages: profiling, optimization, and hierarchical training.

At the \textbf{profiling stage}, HierTrain performs two initialization steps:
(i) profiling the average execution time of different model layers in the
device, edge and cloud workers, respectively; 
(ii) profiling the size of output for each layer in the model. We conduct the profiling by measuring these values in run-time for multiple times, and calculating their mean values.

At the \textbf{optimization stage}, the hierarchical training optimizer selects the best DNN model partition points and determines the number of training samples for the
workers of edge device, edge server and cloud center, respectively. This scheduling policy is generated by the optimization algorithm introduced in Section V.
 The optimization algorithm minimizes the DNN training time with respect to five decision variables $m_s, m_l, b_o, b_s, b_l$ ($m_s, m_l$ represent partition points, $b_o, b_s, b_l$ represent the number of samples processed on each worker, which will be
defined in Section IV). It depends on the following inputs: (i) the profiled average execution time of different model layers in the three workers; (ii) the profiled size of output for each layer in the model; (iii) the available bandwidth between the edge device and the edge server, and that between the edge server and the cloud center.

At the \textbf{hierarchical training stage}, the edge device first sends the delegated data samples to the edge server and the cloud center according to the scheduling policy given in the optimization stage. Once having the needed data samples at hand, the edge device, the edge server and the cloud center start their scheduled training tasks (i.e., the assigned model training modules) immediately, and perform collaborative model training in a hierarchical manner.

Note that the hierarchical training stage depicted in Fig. \ref{fig:systems} only shows one possible scheduling policy, in which the cloud center trains the full model while the edge server and the edge device only train parts of the
model. This scheduling policy is suitable for the scenario that the bandwidth between edge device and cloud center is in a good condition. However, when the network bandwidth becomes the bottleneck, the scheduling policy may choose the
edge server or the edge device to train the full model. In the next section, we will elaborate on how the data samples and the model layers are partitioned.

\section{Problem Statement of Policy Scheduling}
\subsection{Training Tasks in HierTrain}
\begin{figure}[!t]
 \centering
 \includegraphics[scale=0.6]{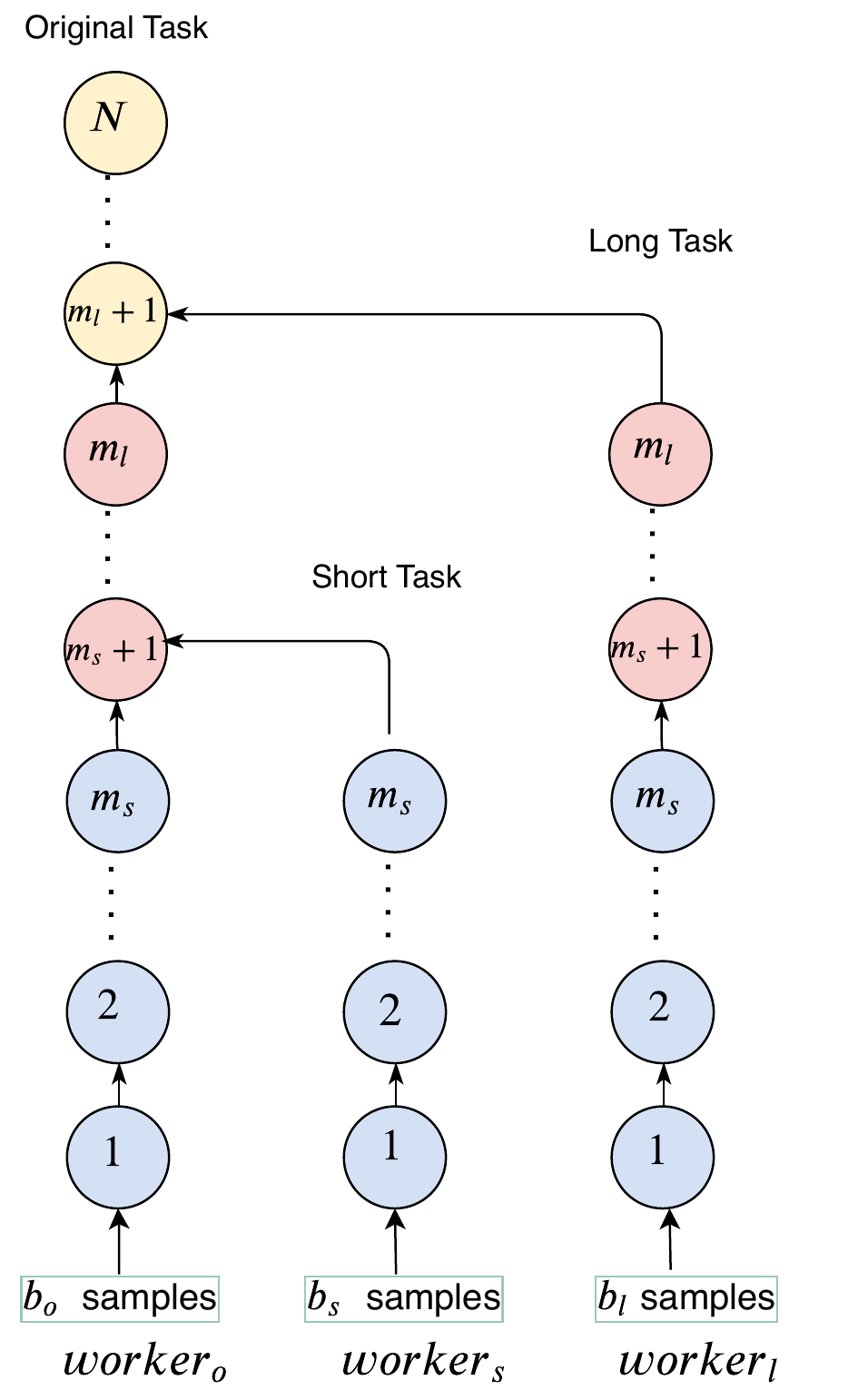}
 \caption{$worker_o$, $worker_s$, $worker_l$ use $b_o$, $b_s$, $b_l$ data samples as
   inputs, respectively. Layer $1$ to layer $m_s$ are executed in parallel over the three workers,
   layer $m_s+1$ to layer $m_l$ are executed in parallel over $worker_o$ and
   $worker_l$, and the rest of layers are executed on $worker_o$.}
 \label{fig:TP}
\end{figure}

We consider that a DNN is stacked by a sequence of distinct layers, and the output of one layer feeds into the input of the next layer. Our goal is to reduce the overall training time in the mobile-edge-cloud environment. Towards this end, we first define three types of training tasks, depicted in Fig. \ref{fig:TP} and explained as follows.

\noindent \texttt{TASK O (Original Task):} Training the full DNN with $b_o$ data samples.

\noindent \texttt{TASK S (Short Task):} Training $m_s$ consecutive layers from layer 1 to layer $m_s$ with $b_s$ data samples.

\noindent \texttt{TASK L (Long Task):} Training $m_l$ consecutive layers from layer 1 to layer $m_l$ with $b_l$ data samples.

\noindent Here $m_s$ and $m_l$ are positive integers, and we assume $m_s \leq m_l \leq N$ ($N$ is the total number of layers in the DNN model).

The key motivations of defining the three task types above are as follows. On one hand, only \texttt{TASK O} contains the most backend layers (e.g., fully
connected layers in many DNNs) that typically have the majority of the parameters, and this helps to reduce the communication overheads for parameter
exchange across different tasks. On the other hand, \texttt{TASK O}, \texttt{L} and \texttt{S} all contain the frontend layers (e.g., convolution layers in many DNNs) that are often computationally intensive, and this also helps to exploit the computing resources of different workers in parallel to accelerate the DNN training. Furthermore, we have the flexibility to optimize the computing workloads of different tasks by varying their input data sample sizes.

In the following, we denote the workers that execute \texttt{TASK O}, \texttt{TASK S} and \texttt{TASK L} as $worker_o$, $worker_s$ and $worker_l$, respectively. We also denote the profiling values ${L}^{f}_{j,i}$, ${L}^{b}_{j,i}$, ${L}^{u}_{j,i}$ and $MP_i$, $MO_i$. Their meanings are shown in Table \ref{tab:notation}.

By defining the three task structures, we have rich flexibility in optimizing the training workloads across the edge device, the edge server and the cloud center by tuning the sizes of their data samples and assigned model layers, tailored to their computation resources and network conditions.

\begin{table}[t!]
\renewcommand{\arraystretch}{1.3}
 \caption{LIST OF NOTATIONS}
  \begin{center}
  \begin{tabular}{|c|c|}
  \hline
    \bfseries Parameter     & \bfseries Description     \\
    \hline
    \hline
    ${L}^{f}_{j,i}$   & forward time to handle $1$ sample for layer $i$ on $worker_j$ \\
    \hline
    ${L}^{b}_{j,i}$   & backward time to handle $1$ sample for layer $i$ on $worker_j$ \\
    \hline
    ${L}^{u}_{j,i}$   & weight update time for layer $i$ on $worker_j$ \\
    \hline
    $MP_i$            & number of parameters in layer $i$ \\
    \hline
    $MO_i$            & output size of layer $i$ in forward phase \\
    \hline
  \end{tabular}
  \label{tab:notation}
   \end{center}
\end{table}

\subsection{Training Procedure in HierTrain}
Based on the above-defined three tasks, we elaborate on the training procedure in HierTrain as follows. First, the scheduling policy determines how to assign the model layers and the data samples to the three workers. Second, the edge device initiates the training procedure and sends the partitioned data samples to the edge server and the cloud center. Last, the following three phases are executed iteratively.

\textit{1) Forward:} $worker_s$ executes the forward phase (i.e., inference through the DNN model to obtain the current model loss) over the assigned layers, using a mini-batch $b_s$ of data samples. Once completing the forward phase over
the assigned layers, $worker_s$ sends the output to $worker_o$. Then, $worker_o$ proceeds to execute the forward phase over the rest of layers. $worker_l$ acts the same as $worker_s$, using a mini-batch $b_l$ of data samples. $worker_o$
also executes the forward phase, but over all the layers and using a mini-batch $b_o$ of data samples. When the forward phase ends, $worker_o$ collects the model losses from $B=b_s+b_l+b_o$ data samples.

\textit{2) Backward:} For each data sample, $worker_o$ starts the backward phase (i.e., back-propagation using the loss to obtain the stochastic gradients) from the last layer of the DNN. If the data sample belongs to $worker_o$, then $worker_o$ executes the full backward phase. If the data sample belongs to $worker_l$, then $worker_o$ sends the intermediate results to $worker_l$ upon reaching layer $m_l+1$, and $worker_l$ proceeds to execute the backward phase over the rest of layers. The same rule applies to $worker_s$, except that $worker_o$ sends the intermediate results to $worker_s$ upon reaching layer $m_s+1$. When the backward phase ends, every worker obtains the stochastic gradients of the assigned layers.

\textit{3) Weight Update:} $worker_l$ and $worker_s$ send the computed stochastic gradients to $worker_o$. Then $worker_o$ averages the stochastic gradients layer-wise, and sends the averaged stochastic gradients to $worker_l$ and $worker_s$ according to the layers assigned to them. With these stochastic gradients, the three workers update the weights of their assigned layers independently.

\subsection{Formulation of Minimizing Training Time}

The core of HierTrain is a scheduling policy that determines how the model layers and the data samples are assigned to the three workers. The goal is to minimize the training time, which is determined by the computation and communication latencies. To analyze these two quantities, we assume that the DNN has $N$ layers and the size of each data sample is $Q$ bits.

\textit{1) Computation Latency:} Recall that the DNN training procedure is divided into three phases: forward, backward and weight update. In the forward and backward phases, the amount of computation is proportional to the number of processed data samples \cite{devarakonda2017adabatch}. We denote $T_{j, i, b, forward}$ and $T_{j, i, b, backward}$ as the computation latencies of executing layer $i$ on $worker_j$ with $b$ input data samples in the forward and backward phases, respectively. Here $j \in \{o, s, l\}$, $i \in \{1, 2, \cdots, N\}$, and $b \in \{b_o, b_o+b_l, b_o+b_l+b_s\}$. Then we have
\begin{IEEEeqnarray}{rCl}
     &T_{j,i,b,forward}  = &  b {L}_{j,i}^{f}, \\
     &T_{j,i,b,backward} = &  b {L}_{j,i}^{b}.
\end{IEEEeqnarray}
The computation latency $T_{j, update}$ of the weight update phase on a worker $j\in\{o,s,l\}$ is the summation of the execution time over the involved layers, given by
\begin{IEEEeqnarray}{C} \label{eq:update}
      T_{j, update} =   \overset{m_j}{\underset{i=1}{\sum}}{L}_{j,i}^{u}.
\end{IEEEeqnarray}

\textit{2) Communication Latency:} The workers are bidirectionally connected with each other. For example, the edge device and the edge server are connected with the high-speed wireless local-area-network (WLAN) link, while the edge server and the cloud center are connected with the bandwidth-limited wide-area-network (WAN) link. Let $B_{o,s}$ denote the bandwidth between $worker_o$ and $worker_s$, $B_{o,l}$ the bandwidth between $worker_o$ and $worker_l$, $B_{s,l}$ the bandwidth between $worker_s$ and $worker_l$. The communication latency is the ratio of the transferred data size and the bandwidth between two workers, as
\begin{IEEEeqnarray}{rCl} \label{eq:transmit}
   T_{communication} = \frac{DataSize}{Bandwidth}.
\end{IEEEeqnarray}
\textit{3) Training Time:} As depicted in Fig. \ref{fig:TP}, $worker_o$, $worker_s$, $worker_l$ use $b_o$, $b_s$, $b_l$ data samples as inputs, respectively. Layer $1$ to layer $m_s$ are executed in parallel over the three workers, layer $m_s+1$ to layer $m_l$ are executed in parallel over $worker_o$ and $worker_l$, and the rest of layers are executed on $worker_o$. Below we calculate the training time, beginning with those in the forward and backward phases.

Denote $T_{forward}^{1}$ and $T_{backward}^{1}$ as the latencies of executing layers between $1$ and $m_s$ over the three workers in the forward and backward phases, respectively, given by
\begin{IEEEeqnarray}{rCl}
  &T_{forward}^{1} = \max \{ & T_{o,input} + \overset{m_s}{\underset{i=1}{\sum}}{T_{o,
      i, b_o, forward}}, \nonumber \label{eq:1} \\
  & &T_{s,input} + \overset{m_s}{\underset{i=1}{\sum}}{T_{s, i, b_s, forward}} +
  T_{s, output},
  \nonumber  \\
  && T_{l, input} + \overset{m_s}{\underset{i=1}{\sum}}{T_{l, i, b_l, forward}}
  \}, \\
  &T_{backward}^{1} = \max & \{ \overset{m_s}{\underset{i=1}{\sum}}{T_{o,
      i, b_o, backward}}, \nonumber \\
  & &\overset{m_s}{\underset{i=1}{\sum}}{T_{s, i, b_s, backward}} + T_{s, grad},
  \nonumber \\
    &&  \overset{m_s}{\underset{i=1}{\sum}}{T_{l, i, b_l, backward}}
    \}.
\end{IEEEeqnarray}
Here $T_{j, input}$ is the communication latency of $worker_j$ to receive $b_j$ data samples, $j \in \{o, s, l\}$. We use \eqref{eq:transmit} to calculate $T_{j, input}$, where $DataSize = b_j \times Q$ and $Bandwidth$ is the bandwidth between the edge device and $worker_j$. $T_{s, output}$ represents the communication latency of $worker_s$ to transmit its forward output to $worker_o$. Recall that $MO_{m_s}$ is the output size of layer $m_s$ in the forward phase for one data sample, $b_s$ is the number of data samples of $worker_s$, and $B_{o,s}$ is the bandwidth between $worker_o$ and $worker_s$. Then according to \eqref{eq:transmit}, $T_{s, output}=\frac{b_s \times MO_{m_s}}{B_{o, s}}$. $T_{s, grad}$ represents the communication latency of
$worker_o$ to send the intermediate results to $worker_s$ in the backward phase. The size of the intermediate results is equal to the output data of layer $m_s$ in forward phase. Thus, $T_{s, grad} = T_{s, output}$.

Denote $T_{forward}^{2}$ and $T_{backward}^{2}$ as the latencies of executing layers between $m_s+1$ and $m_l$ over $worker_o$ and $worker_l$ in the forward and backward phases, respectively, given by
\begin{IEEEeqnarray}{rCl}
  &T_{forward}^{2} = \max \{& \overset{m_l}{\underset{i=m_s+1}{\sum}}{T_{o, i,
       b_o+b_s, forward}} ,
   \nonumber \label{eq:2}  \\
  && \overset{m_l}{\underset{i=m_s+1}{\sum}}{T_{l, i,b_l, forward}}+ T_{l, output}
  \}, \\
  &T_{backward}^{2} = \max & \{\overset{m_l}{\underset{i=m_s+1}{\sum}}{T_{o, i,
      b_o+b_s, backward}},
  \nonumber  \\
     &&\overset{m_l}{\underset{i=m_s+1}{\sum}}{T_{l, i, b_l, backward}} + T_{l, grad}
  \}.
\end{IEEEeqnarray}
Here $T_{l, output}$ is the communication latency of $worker_l$ to transmit its forward output to $worker_o$, given by $T_{l, output} = \frac{b_l \times MO_{m_l}}{B_{o,l}}$.
$T_{l, grad}$ represents the communication latency of $worker_o$ to send the intermediate results to $worker_l$ in the backward phase, it is equal to $T_{l, output}$ .

Denote $T_{forward}^{3}$ and $T_{backward}^{3}$ as the latencies of executing layers between $m_l+1$ and $N$ over $worker_o$ in the forward and backward phases, respectively, given by
\begin{IEEEeqnarray}{rCl}
  &T_{forward}^{3} = & \overset{N}{\underset{i=m_l+1}{\sum}}{T_{o, i,
       b_o+b_s+b_l, forward}},  \\
  &T_{backward}^{3} = & \overset{N}{\underset{i=m_l+1}{\sum}}{T_{o, i,
      b_o+b_s+b_l, backward}}.
\end{IEEEeqnarray}

Now we consider the training time in the weight update phase. After the backward phase finishes, $worker_l$ and $worker_s$ send the stochastic gradients to $worker_o$. Then $worker_o$ sends the averaged stochastic gradients to $worker_l$ and $worker_s$ according to the layers assigned to them, and the three workers update the weights of their assigned layers. The total time cost in the weight update phase is denoted as $T_{update}$, given by
\begin{IEEEeqnarray}{rCl}
  &T_{update} = & \max \{ T_{o, update}, T_{s,update}, T_{l, update} \} \nonumber \\
  && + \max \{ T_{s, weightgrad}, T_{l, weightgrad} \}. \label{Tl}
\end{IEEEeqnarray}
Here $T_{j, update}$ is the computation latency of the weight update phase on $worker_j$, $j \in \{o, s, l\}$, as defined in \eqref{eq:update}. $T_{s, weightgrad}$ and $T_{l, weightgrad}$ represent the communication latencies of $worker_s$ and $worker_l$ to send the stochastic gradients to and receive the updated weights from $worker_o$, respectively. For layer $i$, the sizes of the stochastic gradients and the updated weights are both $MP_i$. Therefore, we have $T_{s, weightgrad} = \frac{2}{B_{o,s}} \sum_{i=1}^{m_s} MP_i$ and $T_{l, weightgrad} = \frac{2}{B_{o,l}}\sum_{i=1}^{m_l} MP_i$.

\textit{4) Minimization of Training Time:} Therefore, the time of training the DNN for one iteration, including both computation and computation, is given by
\begin{IEEEeqnarray}{rCl}
  &T_{total} =  \overset{3}{\underset{k=1}{\sum}}(T_{forward}^{k} + T_{backward}^{k}) +  T_{update},     \label{eq:total}
\end{IEEEeqnarray}
in which the number of used data samples is
\begin{IEEEeqnarray}{C}
    B = b_o + b_s + b_l. \label{consB}
\end{IEEEeqnarray}
Here $B$ is the predefined batch size, while $b_o$, $b_s$ and $b_l$ are decision variables.

The number of layers $m_s$ and $m_l$ for \texttt{TASK S} and \texttt{TASK L} are also decision variables. It is possible in some scenarios that $m_s$ or $m_l$ can equal to 0, meaning that $worker_s$ or $worker_l$ will not participate in the DNN training procedure. For these scenarios, we do not assign any data samples to $worker_s$ or $worker_l$, such that $b_s = 0$ or $b_l =0$. To characterize these connections, we introduce constraints
\begin{IEEEeqnarray}{rCl}
0 \leq & b_s  &\leq m_s B, \label{consbsms} \\
0 \leq & b_l &\le m_l B. \label{consblml}
\end{IEEEeqnarray}
When $m_s=0$ or $m_l=0$, \eqref{consbsms} or \eqref{consblml} ensures that $b_s = 0$ or $b_l =0$. Otherwise, if $m_s$ or $m_l$ is any positive integer, \eqref{consbsms} or \eqref{consblml} automatically satisfies due to \eqref{consB}.

In summary, when $worker_s$, $worker_l$ and $worker_o$ have been fixed, to minimize the training time, HierTrain solves the following optimization problem
\begin{IEEEeqnarray}{rCl}
\mathcal{P}_1:  \ \  &  \underset{\{b_o,b_s,b_l,m_s,m_l \}}{minimize} & ~ T_{total} \\ \nonumber
\\ & s.t.   & ~ b_o + b_s + b_l = B,
   \\ && ~ 0 \leq b_s \leq m_s B,
   \\ && ~ 0 \leq b_l \leq m_l B,
\end{IEEEeqnarray}
where the decision variables $b_o$, $b_s$, $b_l$, $m_s$, $m_l$ are all nonnegative integers. Since there are 6 possible mappings between $worker_s$, $worker_l$, $worker_o$ and the edge device, the edge server, the cloud center, we can enumerate all the mappings, calculate the optimal scheduling policy \{$b_o$, $b_s$, $b_l$, $m_s$, $m_l$\} for each mapping, and then find the global optimal scheduling policy. The next section gives details of the proposed algorithm.

\begin{algorithm}[!t]
  \caption{HierTrain Algorithm}
  \label{APDNN}
  \begin{algorithmic}[1] 
    \State {\textbf{Input}}:
      \begin{enumerate}
      \item{${L}^{f}_{k,i},{L}^{b}_{k,i},{L}^{u}_{k,i}, k \in \{d, e,
          c\}$: profiling values of device, edge, cloud}
        \item{$BW_{de}, BW_{ec}$: bandwidth of device-edge  and edge-cloud}
    \item{$MP_i$:  layer $i$ parameters data size}
    \item{$MO_i$: layer $i$ output data size}
    \end{enumerate}
    \textbf{Output}:
    {optimal solution \{$m_{s}^*$, $m_{l}^*$, $b_{o}^*$, $b_{s}^*$, $b_{l}^*$\}}
    \State {\textbf{Initialization}:
    {$T_{total, minimum}$ = MAX} \Comment{$MAX$ is an infinite number} }
    \For{$\{node_o, node_s, node_l\} \leftarrow permutation\{d, e, c\}$}
    \Comment{ map \{devcie, edge, cloud\} to \{$node_o, node_s, node_l$\}}
       \For{$m_s = 0 \to N$}
          \For{$m_l = m_s \to N$}
             \State {Solve  problem $\mathcal{P}_1$ to get $\{b_o, b_s, b_l\}$}
             \State {\{$b_o, b_s, b_l$\} $\leftarrow Round(b_o, b_s, b_l)$}
             \Comment {rounding $b_o,b_s,b_l$ to integers}
             \State {Calculate $T_{total}$ according to \eqref{eq:total}}
             \If{$T_{total} < T_{total, minimum}$}
             \State {
                  \{$m_s^*, m_l^*,b_o^*, b_s^*, b_l^*$\}  =
                 \{ $m_s, m_l, b_o, b_s, b_l$\}
             }
                \State {$T_{total, minimum} = T_{total}$}
             \EndIf
          \EndFor
       \EndFor
    \EndFor
  \end{algorithmic}
  Return: \{$m_s^*, m_l^*,b_o^*, b_s^*, b_l^* $\}
\end{algorithm}

\section{Optimization of Scheduling Policy}

Note that even when $worker_s$, $worker_l$ and $worker_o$ have been fixed, solving $\mathcal{P}_1$ is still challenging because: (i) in the objective $T_{total}$, the terms of $T_{update}$, $T_{forward}^{k}$ and $T_{backward}^{k}$, where $k = 1, 2, 3$, all contain summations with the numbers of summands determined by $m_s$ and $m_l$; (ii) the decision variables $b_o$, $b_s$, $b_l$, $m_s$, $m_l$ are all integers.

To address the first challenge, we observe that when $m_s$ and $m_l$ are fixed, $\mathcal{P}_1$ will become a standard integer linear programming (ILP) problem and is relatively easier to solve. Motivated by this observation, we enumerate
the values of $m_s$ and $m_l$, solve the resulting ILP problems, and then find the best one among the ILP solutions. This enumeration is feasible because the numbers of layers $m_s$ and $m_l$ are often modest in practice (such as
AlexNet: 8 layers, VGG-16: 16 layers, GoogleNet: 22 layers, MobileNet: 28 layers).

To address the second challenge, for each ILP problem, we relax the integer variables to real ones, solve the relaxed linear programming (LP) problem, and then round the solution to integers. To be specific, the relaxed LP problem can be efficiently solved with CPLEX, Gurobi or CVXPY. Although these optimization solvers can solve ILP problem directly, we choose to convert the ILP problem to LP problem the reason is that these solvers solve LP problem are much faster than solve ILP
problem. Further, the \textbf{rounding operation} works as follows. Given a real solution $(b_o, b_s, b_l)$ of the relaxed LP problem, we divide them into integer parts $int(b_j)$ and fraction parts $frac(b_j)$, $j \in \{o, s, l\}$, and then sort the fraction parts in a descending order. For $b_j$ with the largest fraction part, we let $b_j^* = int(b_j)+1$, while for the other two $b_j$, $b_j^* = int(b_j)$. If $b_o^* + b_s^* + b_l^* = B$ is satisfied, then the rounding operation ends. Otherwise, for the two $b_j$ with the largest fraction parts, we let $b_j^* = int(b_j)+1$, while for the other $b_j$, $b_j^* = int(b_j)$. The constraint $b_o + b_s + b_l = B$ can be satisfied after at most two steps.

\begin{figure}[h!]
 \centering
 \includegraphics[width=3.5in]{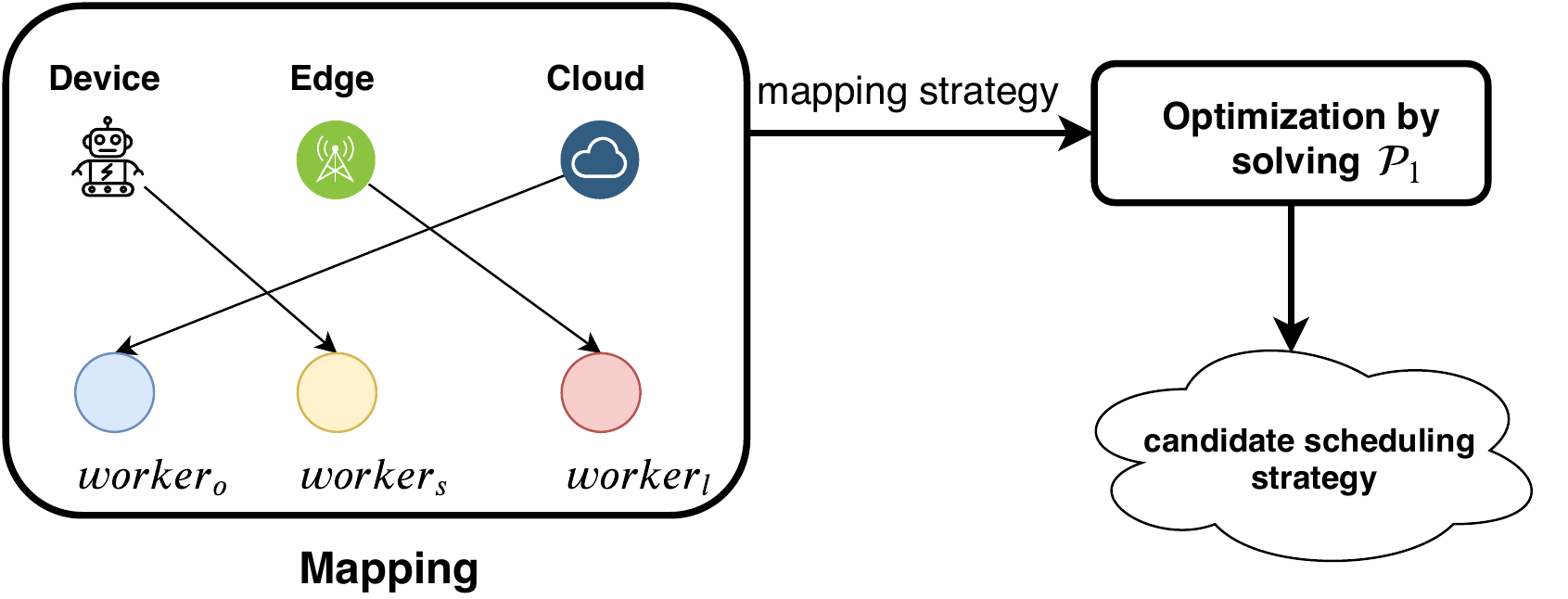}
 \caption{Each mapping strategy corresponds to a candidate optimal scheduling policy.}
 \label{algo}
\end{figure}

So far, we have solved $\mathcal{P}_1$ given that $worker_s$, $worker_l$ and $worker_o$ have been fixed. In order to deploy the DNN training task over the device-edge-cloud environment, we still need to find the best mapping strategy between the device, edge and cloud workers and $worker_o$, $worker_s$ and $worker_l$. As illustrated in Fig. \ref{algo}, since the overall number of mappings is only 6, we can enumerate all the mapping, find a candidate optimal scheduling policy for each mapping, and then choose the best mapping strategy with the minimum training time. The algorithm is outlined in \textbf{Algorithm} \ref{APDNN}.

As  shown in the Table \ref{tab:time}, in order to verify the efficiency of our algorithm, we list the algorithm running time  based on some common deep networks configuration. All results are obtained on a desktop computer equipped with an Intel Core(TM) i7-6700 3.4 GHz with 8 GB RAM runing Linux. We use python as programming language and CPLEX as optimization problem solver. From Table \ref{tab:time}, we see than the proposed algorithm runs very fast, and in practice its running time can be ignored compared with the long DNN training time.

\begin{table}[!t]
\renewcommand{\arraystretch}{1.3}
\caption{Algorithm Running Time}
 \begin{center}
 \begin{tabular}{|c|c|c|c|c|c|}
 \hline
   \bfseries LeNet     & \bfseries AlexNet  & \bfseries VGG-16 & \bfseries VGG-19  & \bfseries googLeNet  & \bfseries ResNet-34 \\
   \hline
   0.52s  & 1.48s & 3s & 4s & 5.3s & 12s \\
   \hline
 \end{tabular}
 \label{tab:time}
  \end{center}
\end{table}

\section{Evaluation}
\subsection{Dataset and Models}
We evaluate HierTrain by training two well-known CNNs for image classification tasks. The first CNN is LeNet-5 \cite{lecun1998gradient}, and we train it with the CIFAR-10 dataset \cite{krizhevsky2009learning}. CIFAR-10 contains 50,000 training images and 10,000 testing images, each of which has 10 labels. The second CNN is AlexNet \cite{krizhevsky2012imagenet}, which is more complicated than LeNet-5. We train AlexNet on the tiny ImageNet dataset. The tiny ImageNet dataset has 200 classes, while each class has 500 training images, 50 validation images, and 50 testing images.

\subsection{Experimental Setup}
We use a Raspberry Pi 3 tiny computer to act as an edge device. The Raspberry Pi 3 has a quad-core ARM processor at 1.2 GHz with 1 GB of RAM. We use an Intel NUC, a small but powerful mini PC which is
equipped with a four Intel Cores (TM) i3-7100U with 8 GB of RAM, to emulate the edge server. Unless specifically indicated, we only use one core of the edge server in our experiments. This is to simulate the application scenarios where the edge server has to serve multiple edge devices and each edge device cannot occupy all the computation resource of the edge server. The cloud center is a Dell Precision T5820 Tower workstation with 16 Intel Xeon processor at 3.7 GHz and with 30 GB of RAM, and equipped with NVIDIA GPU GeForce GTX 1080 Ti. The computation capability of the cloud center is one order magnitude higher than those of the edge device and the edge server. All the three workers run the Ubuntu system, and we use Linux Traffic Control on them to emulate constrained network bandwidths.

There are many existing open-source platforms for training CNNs, such as TensorFlow \cite{abadi2016tensorflow}, Theano \cite{bergstra2010theano}, MXNet \cite{chen2015mxnet}, PyTorch \cite{paszke2017automatic}, and Chainer \cite{tokui2015chainer}. Among them we choose Chainer because it is flexible and able to leverage dynamic computation graphs, which facilitates the application of the proposed \textit{hybrid parallelism} method.

\subsection{Baselines}
To elucidate the performance of the proposed HierTrain framework, we consider the following baselines in the experimental evaluation.

\textit{1) All-Edge:} The edge device transmits all the training data samples to the edge server, and the edge server completes the DNN training.

\textit{2) All-Cloud:} The edge device transmits all the training data samples to the cloud center, and the cloud center completes the DNN training.

\textit{3) JointDNN \cite{eshratifar2018jointdnn}:} The edge device and the cloud center jointly train the DNNs.

\textit{4) JointDNN+:} We extend JointDNN to train the DNNs in the mobile-edge-cloud architecture. Following the design of JointDNN, the scheduling in JointDNN+ is by solving a shortest path problem over a graphic model.

\textit{5) JALAD \cite{li2018jalad}:} The edge server and the cloud center jointly train the DNNs. A data compression strategy is applied to reduce the edge-cloud
transmission latency. In our experiments we set the number of bits $c$ used in data compression as 8.

\subsection{Results}

\begin{figure}[t!]
 \centering
 \includegraphics[width=2.4in]{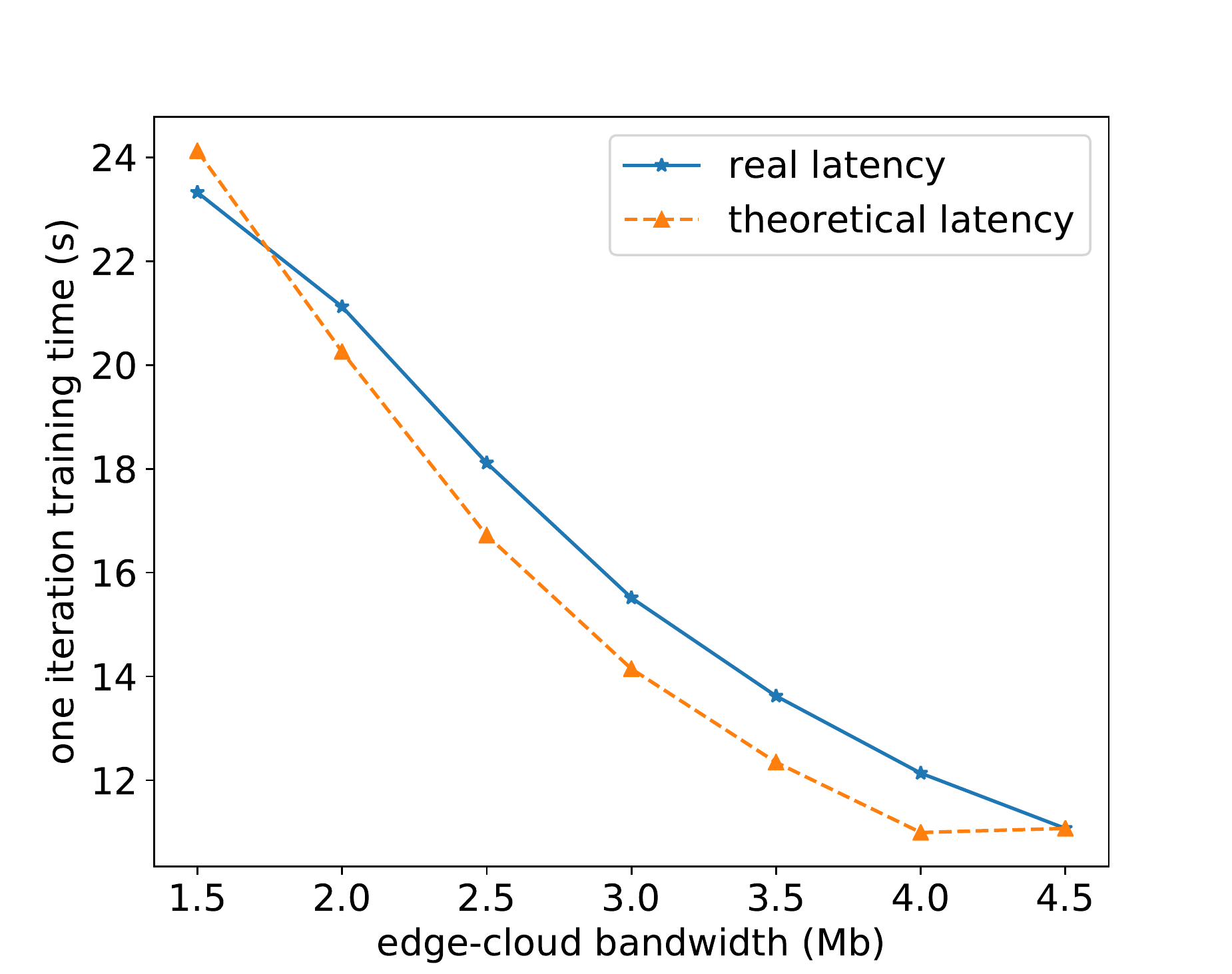}
 \caption{Comparison of real and theoretical latencies of training AlexNet.}
 \label{realvstheory}
\end{figure}

\textit{1) Model Validity:} We first validate the formulated model that captures the execution delay of one iteration in training a DNN. Using the same scheduling policy, we obtain the real latency measured from the experiment and the theoretical latency, both in training AlexNet. As is shown in Fig. \ref{realvstheory}, the real and theoretical latencies highly match.

\begin{figure}[t!]
 \centering
 \includegraphics[width=2.4in]{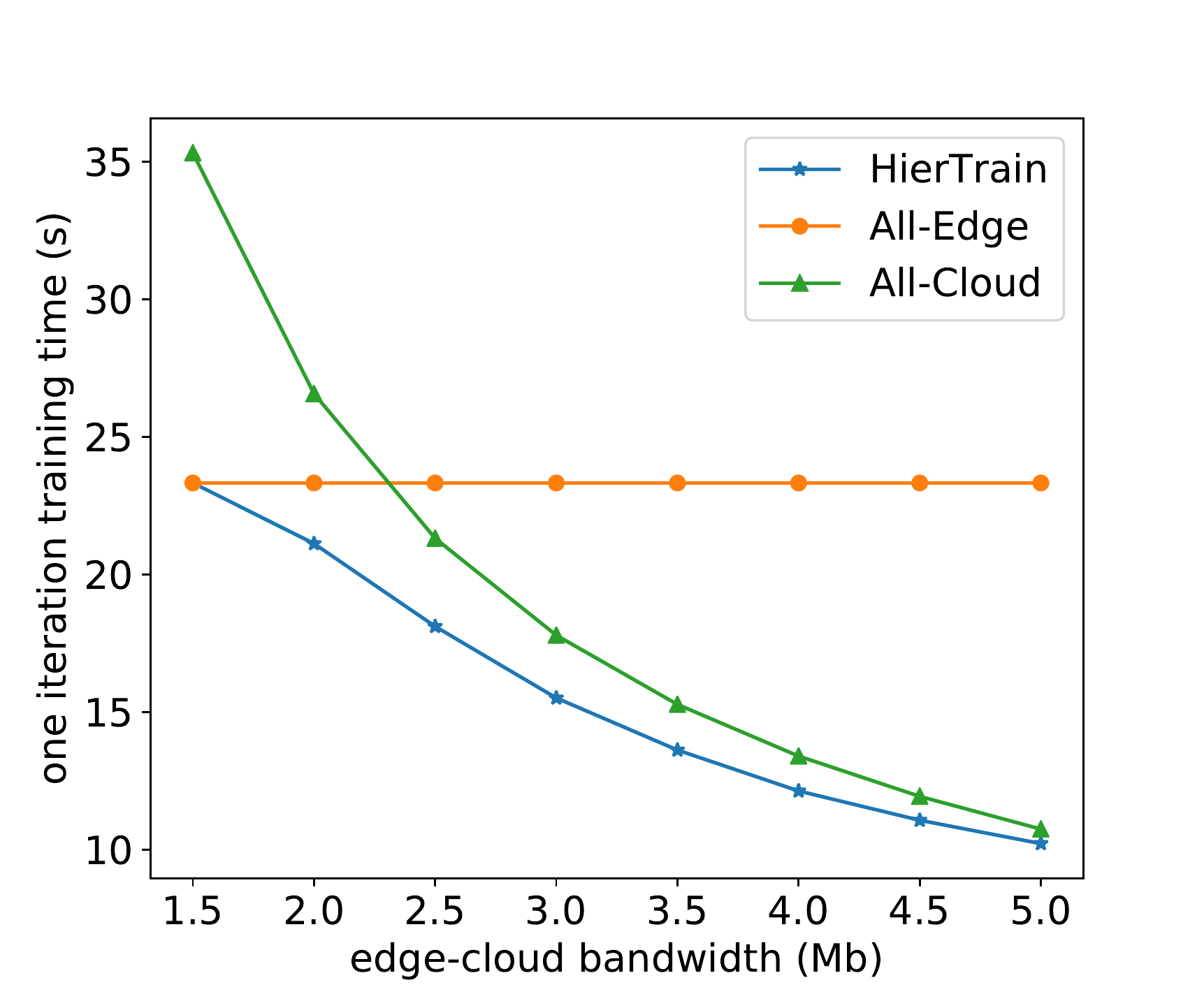}
 \caption{Per-iteration training time of AlexNet for HierTrain, All-Edge and All-Cloud under different bandwidths.}
 \label{alexNet1}
\end{figure}

\begin{figure}[t!]
 \centering
 \includegraphics[width=2.4in]{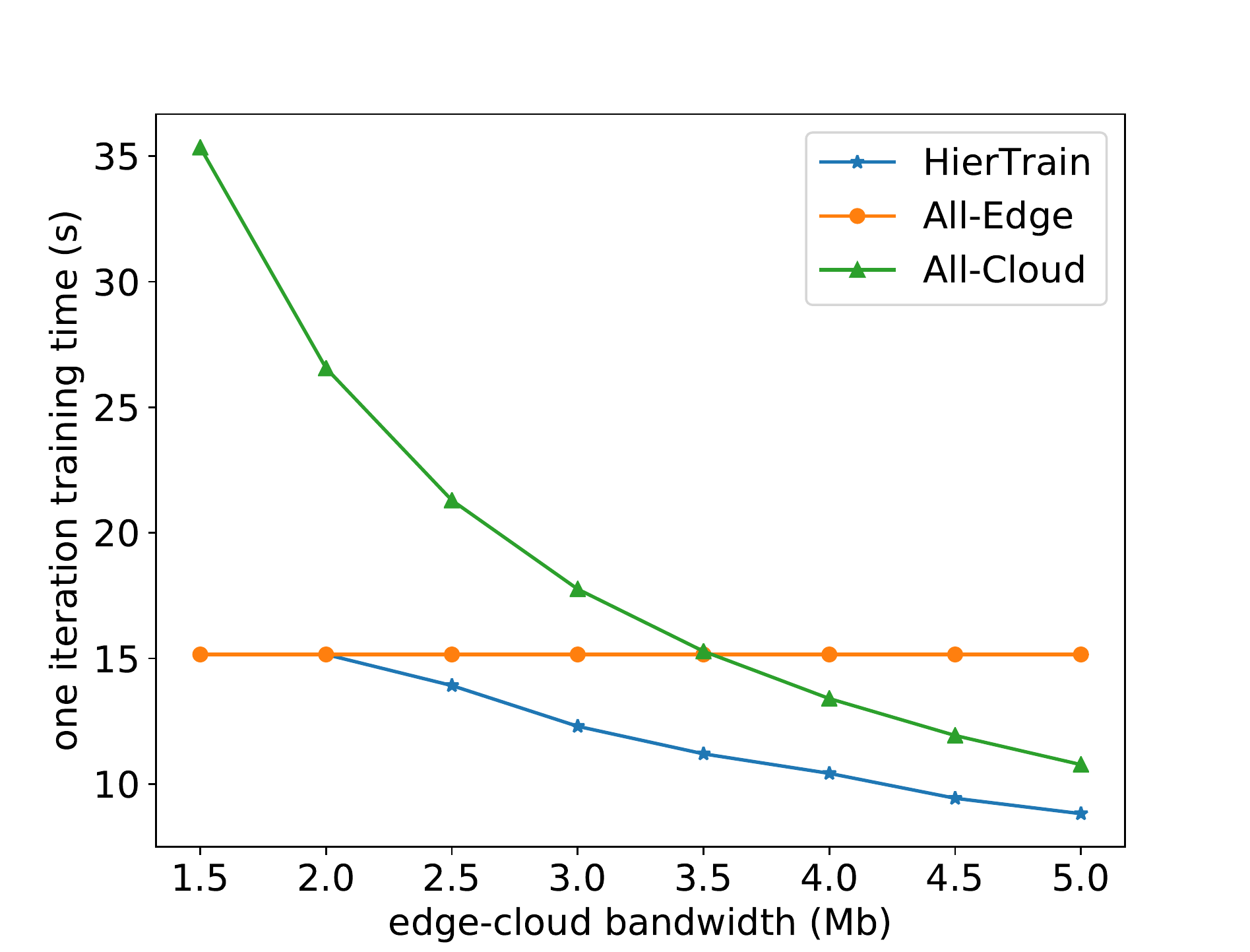}
 \caption{Per-iteration training time of LeNet-5 for HierTrain, All-Edge and All-Cloud under different bandwidths.}
 \label{lenet51}
\end{figure}

\textit{2) Comparison with All-Edge and All-Cloud:} Next we compare HeirTrain with the two baselines, All-Edge and All-Cloud, by fixing the mobile-edge bandwidth to 5 Mbps and varying the edge-cloud bandwidth from 1.5 Mbps to 5 Mbps. Fig. \ref{alexNet1} shows the average per-iteration time to train AlexNet. The time cost of All-Cloud decreases as the edge-cloud bandwidth increases, while that of All-Edge remains unchanged. HeirTrain outperforms, and achieves up to 2.3$\times$ and 4.5$\times$ speedup comparing to
All-Edge and All-Cloud, respectively. Similar observations can be found in training LeNet-5, as depicted in Fig. \ref{lenet51}. HierTrain is the best among the three schemes, achieves up to 1.7$\times$ and 6.9$\times$ speedup comparing to All-Edge and All-Cloud, respectively.

\begin{figure}[t!]
 \centering
 \includegraphics[width=2.4in]{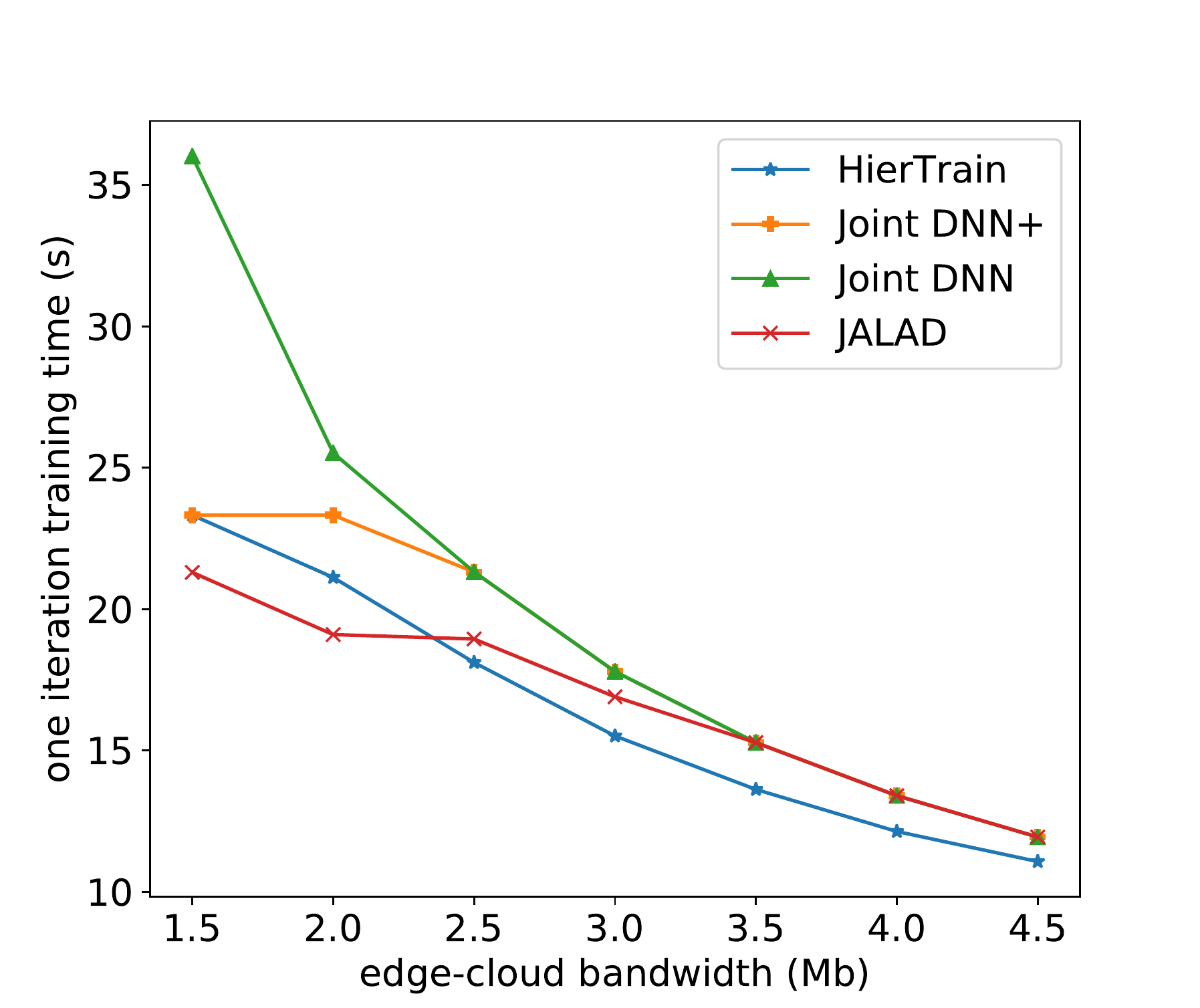}
 \caption{Per-iteration training time of AlexNet for HierTrain, JointDNN, JointDNN+, and JALAD under different bandwidths.}
 \label{alexnet2}
\end{figure}

\begin{figure}[t!]
 \centering
 \includegraphics[width=2.4in]{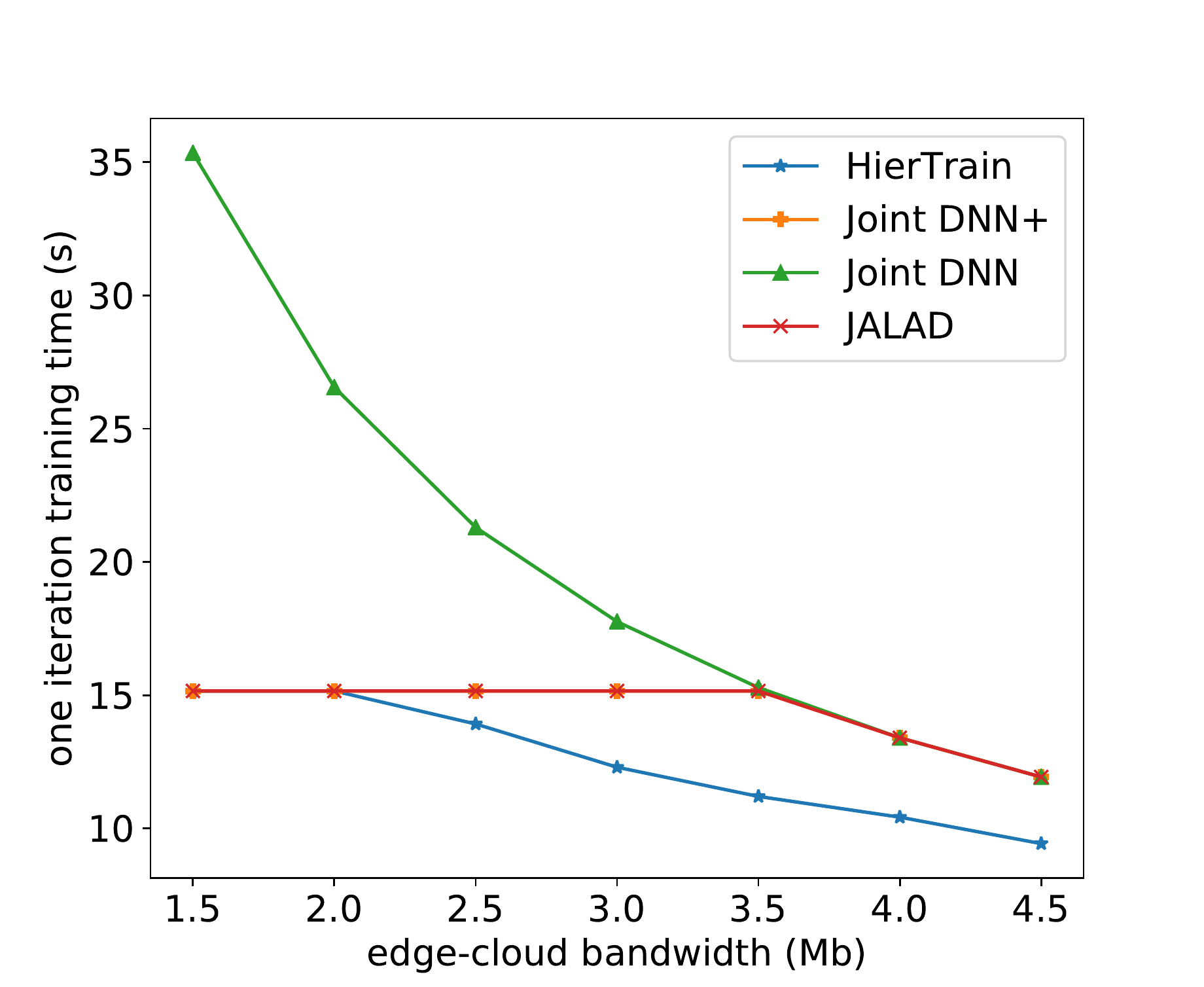}
 \caption{Per-iteration training time of LeNet-5 for HierTrain, JointDNN, JointDNN+, and JALAD under different bandwidths.}
 \label{lenet52}
\end{figure}

\textit{3) Comparison with JointDNN, JointDNN+ and JALAD:} Now we conduct experiments to compared HierTrain with the three baselines: two state-of-the-art methods JointDNN and JALAD, as well as JointDNN+ that extends JointDNN to the mobile-edge-cloud architecture. The results on training AlexNet and LeNet-5 are demonstrated in Fig. \ref{alexnet2} and Fig. \ref{lenet52}, respectively. Observe that HierTrain outperforms both JointDNN and JointDNN+. Among these two baselines, JointDNN+ is better than JointDNN because it can utilize the edge server when the edge-cloud bandwidth is as low as 1.5 Mbps or 2 Mbps. When the edge-cloud bandwidth becomes larger, both JointDNN and JointDNN+ choose to run the training tasks in the cloud center.

Fig. \ref{alexnet2} also compares HierTrain and JALAD in training AlexNet. When the edge-cloud bandwidth ranges from 1.5 Mbps to 2 Mbps, JALAD performs better than HierTrain. The reason is that the data compression strategy of JALAD can largely reduce the amount of transmitted data between the edge server and the cloud center. This makes JALAD advantageous in the low bandwidth condition as the communication time cost is the dominating factor in the overall delay. However, when the bandwidth increases, the benefit from reducing communication delay with data compression degrades sharply, and HierTrain outperforms JALAD. In Fig. \ref{lenet52} that shows the experimental results of training LeNet-5, the curves of JALAD and JointDNN+ overlap, because their scheduling policies are the same in the scenario -- they are the same as the All-Edge strategy in the low bandwidth condition and the All-Cloud strategy in the high bandwidth condition.

\begin{figure}[t!]
 \centering
 \includegraphics[width=2.4in]{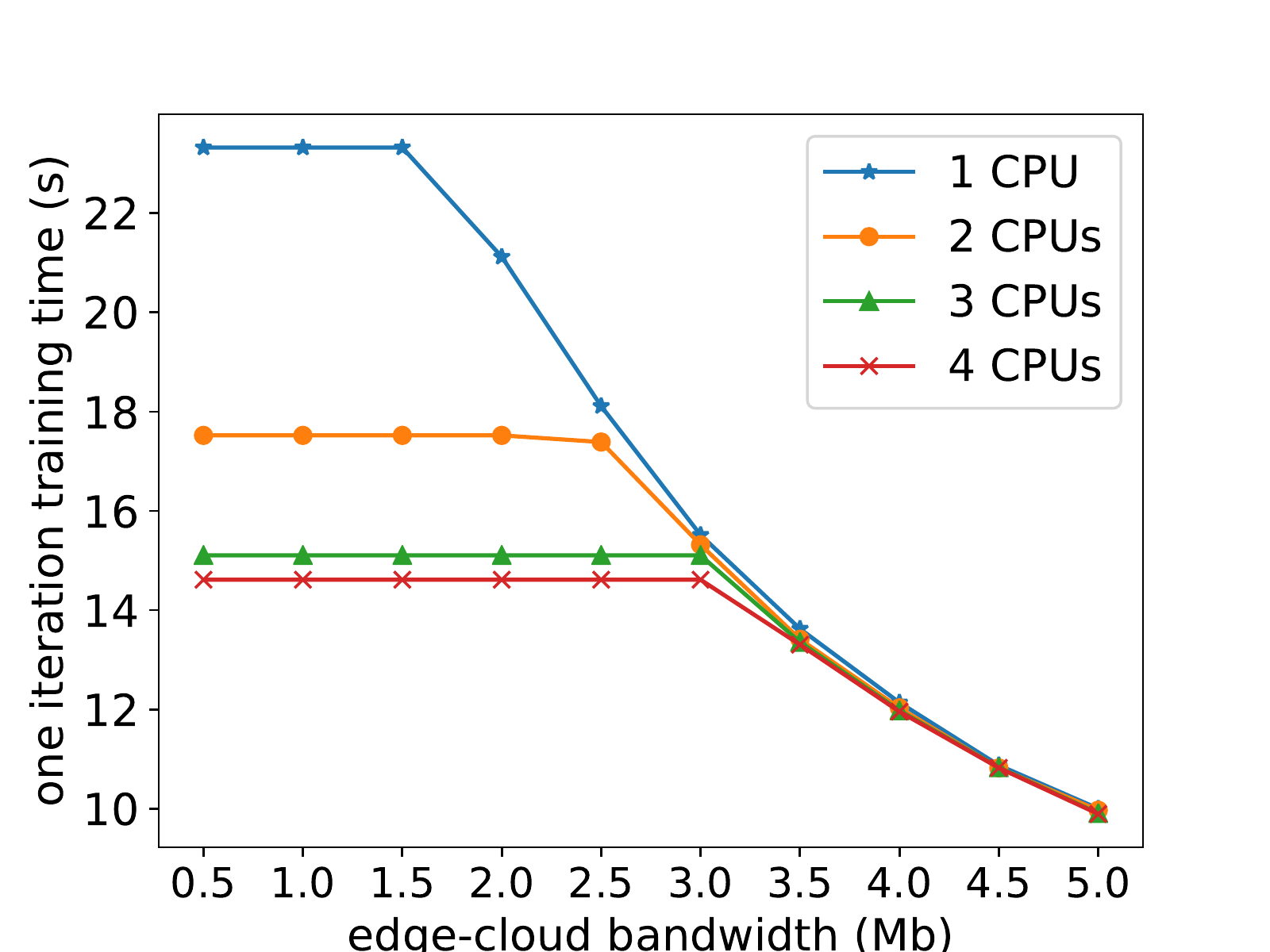}
 \caption{Effect of varying computation capability of edge server in HierTrain under different bandwidths.}
 \label{multicpu}
\end{figure}

\textit{4) Effect of varying edge server resources:} Finally, we investigate the performance of HierTrain when the computation capability at the edge server changes. We consider training AlexNet, while keep the mobile-edge bandwidth as 5 Mbps and the edge-cloud bandwidth as 3.5 Mbps. We use docker to control the CPU cores used in the training process. As shown in Fig. \ref{multicpu}, when the edge-cloud bandwidth is very low ($\leq 1.5$ Mbps), improving the computation capability of the edge server can speedup the training process. This performance gain shrinks when the computation capability of the edge server keeps increasing. To be specific, varying from 1 CPU to 2 CPUs leads to large speedup, while varying from 3 CPUs to 4 CPUs yields insignificant speedup. When the edge-cloud bandwidth is sufficiently large ($\geq 3$ Mbps), the computation capability of the edge server does not influence the overall performance. The reason for this phenomenon is that when the edge-cloud bandwidth is sufficiently large, the optimal policy is training on the cloud.


\section{Related Work}
Due to the attractive features of elasticity in computing power and flexible collaboration, hierarchically distributed computing structures naturally become a popular choice for executing DNN training or inference. Considering the deployment location for DNN, existing approaches can be divided into three classes.

\subsection{Cloud-Based} 
Conventionally, most DNNs are usually deployed on the powerful cloud datacenters \cite{skala2015scalable}. However, this means that a large amount of original data should be uploaded to the cloud,
causing prohibitive communication overhead. In order to improve efficiency, Neurosurgeon \cite{kang2017neurosurgeon} proposed a computation offloading idea in DNNs between the edge device and the cloud server at layer-granularity.
Neurosurgeon explored one suitable partition point of DNN model and the execution starts with edge device and then switches to the cloud, which performs the rest of the computation. \cite{eshratifar2018energy} presented an optimal
scheduling algorithm for collaboratively computation of feed-forward neural networks to achieve maximum performance and energy efficiency. JointDNN\cite{eshratifar2018jointdnn} provided optimization formulations at layer-granularity for forward and backward
propagation in DNNs, which can get the optimal computation scheduling of processing some layers on the edge device and some layers on the cloud server. The limitation of cloud-based approach is that the long WAN latency between
device and cloud.

\subsection{Edge-Based} 
An alternative is to deploy DNN at the edge of network. Li et al.\cite{li2018edge} proposed a collaborative and on-demand DNN co-inference framework which could leverage hybrid computation resources of
device and edge so as to achieve on-demand low-latency. \cite{satyanarayanan2009case} exploited the virtual machine technique to let mobile users utilize nearby server called ``cloudlet'' to speed up service. Gabriel \cite{ha2013just, ha2014towards} is a system that uses ``cloudlet'' for speech and face recognition applications. The focus of these works above is DNN inference at the edge. For the edge learning that considers the DNN training, many existing works target at the fast and cost-efficient
federated learning scheme in order to train a commonly-shared model across multiple devices \cite{lim2019federated}. Along a different line, we consider the fast model learning with respect to a specific end device and leverage a multitude of device-edge-cloud resources to training acceleration.

\subsection{Hierarchy-Based} 
Alternatively, some studies focus on using both central cloud servers and edge servers for the execution. \cite{teerapittayanon2017distributed} proposed a novel distributed DNN framework over distributed computing hierarchies (consisting of cloud, edge, devices), which can allow low-latency classification via early exit. Li et al.\cite{li2018jalad} decoupled the DNN to execute on an edge and the cloud. They not only take into account
latency measurement and raw data quantity between layers, but also take the compression of in-layer data into account. Huang et al.\cite{huang2019deepar} proposed a DeePar framework which can exploit all the available resources from the device, the
edge, and the cloud to improve the overall inference performance. Lin et al.\cite{lin2019cost} proposed a cost-driven offloading strategy based on a self-adaptive particle swarm optimization (PSO) algorithm using the genetic algorithm (GA) operators (PSO-GA) to optimize the system cost during offloading DNN layers over the cloud, edge, and devices.
 
Previous studies above mainly focus on distributed DNN inference. And they follow the scheme of partitioning DNNs into several parts then executing sequentially, which could not fully utilize the computation resources. In our work, we consider accelerating training DNNs in a hierarchical computing paradigm. To this end, we propose the training methodology \textit{hybrid parallelism} which can dynamically adapt the number of parallel execution layers over computing nodes. In addition, different from previous studies we separate computation
overhead not only on layer-granularity but also on sample-granularity.

\section{Conclusions}
In this paper, we study the problem of accelerating the training procedure of DNNs on the device-edge-cloud architecture. To this end, first, we present a novel \textit{hybrid parallelism} method for training DNNs. Secondly, in order get scheduling policy of using \textit{hybrid parallelism} method to train DNNs on the device-edge-cloud environment, we formulate the problem of computation scheduling of training DNNs at layer-granularity and sample-granularity as a minimization optimization programming problem, and solve it to get the scheduling policy. In addition, we test HierTrain in the real hardware and the results show that  it could obviously outperform the naive policy such as all-edge and all-cloud, and also outperform exist prior work like JointDNN and JALAD.

For the future work, we are going to generalize the HierTrain framework to the application scenarios in multi-device and multi-edge environments, in which the federated learning across multi-devices and the device-to-edge association are interesting and challenging.


\bibliographystyle{IEEEtran}
\bibliography{ref}

\begin{thebibliography}{10}
\providecommand{\url}[1]{#1}
\csname url@samestyle\endcsname
\providecommand{\newblock}{\relax}
\providecommand{\bibinfo}[2]{#2}
\providecommand{\BIBentrySTDinterwordspacing}{\spaceskip=0pt\relax}
\providecommand{\BIBentryALTinterwordstretchfactor}{4}
\providecommand{\BIBentryALTinterwordspacing}{\spaceskip=\fontdimen2\font plus
\BIBentryALTinterwordstretchfactor\fontdimen3\font minus
  \fontdimen4\font\relax}
\providecommand{\BIBforeignlanguage}[2]{{%
\expandafter\ifx\csname l@#1\endcsname\relax
\typeout{** WARNING: IEEEtran.bst: No hyphenation pattern has been}%
\typeout{** loaded for the language `#1'. Using the pattern for}%
\typeout{** the default language instead.}%
\else
\language=\csname l@#1\endcsname
\fi
#2}}
\providecommand{\BIBdecl}{\relax}
\BIBdecl

\bibitem{simonyan2014very}
K.~Simonyan and A.~Zisserman, ``Very deep convolutional networks for
  large-scale image recognition,'' \emph{arXiv preprint arXiv:1409.1556}, 2014.

\bibitem{devlin2018bert}
J.~Devlin, M.-W. Chang, K.~Lee, and K.~Toutanova, ``Bert: Pre-training of deep
  bidirectional transformers for language understanding,'' \emph{arXiv preprint
  arXiv:1810.04805}, 2018.

\bibitem{covington2016deep}
P.~Covington, J.~Adams, and E.~Sargin, ``Deep neural networks for youtube
  recommendations,'' in \emph{Proceedings of the 10th ACM conference on
  recommender systems}.\hskip 1em plus 0.5em minus 0.4em\relax ACM, 2016, pp.
  191--198.

\bibitem{zhou2019edge}
Z.~Zhou, X.~Chen, E.~Li, L.~Zeng, K.~Luo, and J.~Zhang, ``Edge intelligence:
  Paving the last mile of artificial intelligence with edge computing,''
  \emph{Proceedings of the IEEE}, vol. 107, no.~8, pp. 1738--1762, 2019.

\bibitem{kumar2019comprehensive}
M.~Kumar, S.~Sharma, A.~Goel, and S.~Singh, ``A comprehensive survey for
  scheduling techniques in cloud computing,'' \emph{Journal of Network and
  Computer Applications}, 2019.

\bibitem{zhang2017poseidon}
H.~Zhang, Z.~Zheng, S.~Xu, W.~Dai, Q.~Ho, X.~Liang, Z.~Hu, J.~Wei, P.~Xie, and
  E.~P. Xing, ``Poseidon: An efficient communication architecture for
  distributed deep learning on $\{$GPU$\}$ clusters,'' in \emph{2017
  $\{$USENIX$\}$ Annual Technical Conference ($\{$USENIX$\}$$\{$ATC$\}$ 17)},
  2017, pp. 181--193.

\bibitem{mathur2018hydra}
V.~Mathur and K.~Chahal, ``Hydra: A peer to peer distributed training \& data
  collection framework,'' \emph{arXiv preprint arXiv:1811.09878}, 2018.

\bibitem{eshratifar2018jointdnn}
A.~E. Eshratifar, M.~S. Abrishami, and M.~Pedram, ``Jointdnn: an efficient
  training and inference engine for intelligent mobile cloud computing
  services,'' \emph{arXiv preprint arXiv:1801.08618}, 2018.

\bibitem{ren2019accelerating}
J.~Ren, G.~Yu, and G.~Ding, ``Accelerating dnn training in wireless federated
  edge learning system,'' \emph{arXiv preprint arXiv:1905.09712}, 2019.

\bibitem{li2019edge}
E.~Li, L.~Zeng, Z.~Zhou, and X.~Chen, ``Edge ai: On-demand accelerating deep
  neural network inference via edge computing,'' \emph{IEEE Transactions on
  Wireless Communications}, 2019.

\bibitem{murshed2019machine}
M.~Murshed, C.~Murphy, D.~Hou, N.~Khan, G.~Ananthanarayanan, and F.~Hussain,
  ``Machine learning at the network edge: A survey,'' \emph{arXiv preprint
  arXiv:1908.00080}, 2019.

\bibitem{bottou2010large}
L.~Bottou, ``Large-scale machine learning with stochastic gradient descent,''
  in \emph{Proceedings of COMPSTAT'2010}.\hskip 1em plus 0.5em minus
  0.4em\relax Springer, 2010, pp. 177--186.

\bibitem{li2018jalad}
H.~Li, C.~Hu, J.~Jiang, Z.~Wang, Y.~Wen, and W.~Zhu, ``Jalad: Joint
  accuracy-and latency-aware deep structure decoupling for edge-cloud
  execution,'' in \emph{2018 IEEE 24th International Conference on Parallel and
  Distributed Systems (ICPADS)}.\hskip 1em plus 0.5em minus 0.4em\relax IEEE,
  2018, pp. 671--678.

\bibitem{goyal2017accurate}
P.~Goyal, P.~Doll{\'a}r, R.~Girshick, P.~Noordhuis, L.~Wesolowski, A.~Kyrola,
  A.~Tulloch, Y.~Jia, and K.~He, ``Accurate, large minibatch sgd: Training
  imagenet in 1 hour,'' \emph{arXiv preprint arXiv:1706.02677}, 2017.

\bibitem{you2017large}
Y.~You, I.~Gitman, and B.~Ginsburg, ``Large batch training of convolutional
  networks,'' \emph{arXiv preprint arXiv:1708.03888}, 2017.

\bibitem{devarakonda2017adabatch}
A.~Devarakonda, M.~Naumov, and M.~Garland, ``Adabatch: adaptive batch sizes for
  training deep neural networks,'' \emph{arXiv preprint arXiv:1712.02029},
  2017.

\bibitem{lecun1998gradient}
Y.~LeCun, L.~Bottou, Y.~Bengio, P.~Haffner \emph{et~al.}, ``Gradient-based
  learning applied to document recognition,'' \emph{Proceedings of the IEEE},
  vol.~86, no.~11, pp. 2278--2324, 1998.

\bibitem{krizhevsky2009learning}
A.~Krizhevsky, G.~Hinton \emph{et~al.}, ``Learning multiple layers of features
  from tiny images,'' Citeseer, Tech. Rep., 2009.

\bibitem{krizhevsky2012imagenet}
A.~Krizhevsky, I.~Sutskever, and G.~E. Hinton, ``Imagenet classification with
  deep convolutional neural networks,'' in \emph{Advances in neural information
  processing systems}, 2012, pp. 1097--1105.

\bibitem{abadi2016tensorflow}
M.~Abadi, P.~Barham, J.~Chen, Z.~Chen, A.~Davis, J.~Dean, M.~Devin,
  S.~Ghemawat, G.~Irving, M.~Isard \emph{et~al.}, ``Tensorflow: A system for
  large-scale machine learning,'' in \emph{12th $\{$USENIX$\}$ Symposium on
  Operating Systems Design and Implementation ($\{$OSDI$\}$ 16)}, 2016, pp.
  265--283.

\bibitem{bergstra2010theano}
J.~Bergstra, O.~Breuleux, F.~Bastien, P.~Lamblin, R.~Pascanu, G.~Desjardins,
  J.~Turian, D.~Warde-Farley, and Y.~Bengio, ``Theano: a cpu and gpu math
  expression compiler,'' in \emph{Proceedings of the Python for scientific
  computing conference (SciPy)}, vol.~4, no.~3.\hskip 1em plus 0.5em minus
  0.4em\relax Austin, TX, 2010.

\bibitem{chen2015mxnet}
T.~Chen, M.~Li, Y.~Li, M.~Lin, N.~Wang, M.~Wang, T.~Xiao, B.~Xu, C.~Zhang, and
  Z.~Zhang, ``Mxnet: A flexible and efficient machine learning library for
  heterogeneous distributed systems,'' \emph{arXiv preprint arXiv:1512.01274},
  2015.

\bibitem{paszke2017automatic}
A.~Paszke, S.~Gross, S.~Chintala, G.~Chanan, E.~Yang, Z.~DeVito, Z.~Lin,
  A.~Desmaison, L.~Antiga, and A.~Lerer, ``Automatic differentiation in
  pytorch,'' 2017.

\bibitem{tokui2015chainer}
S.~Tokui, K.~Oono, S.~Hido, and J.~Clayton, ``Chainer: a next-generation open
  source framework for deep learning,'' in \emph{Proceedings of workshop on
  machine learning systems (LearningSys) in the twenty-ninth annual conference
  on neural information processing systems (NIPS)}, vol.~5, 2015, pp. 1--6.

\bibitem{skala2015scalable}
K.~Skala, D.~Davidovic, E.~Afgan, I.~Sovic, and Z.~Sojat, ``Scalable
  distributed computing hierarchy: Cloud, fog and dew computing,'' \emph{Open
  Journal of Cloud Computing (OJCC)}, vol.~2, no.~1, pp. 16--24, 2015.

\bibitem{kang2017neurosurgeon}
Y.~Kang, J.~Hauswald, C.~Gao, A.~Rovinski, T.~Mudge, J.~Mars, and L.~Tang,
  ``Neurosurgeon: Collaborative intelligence between the cloud and mobile
  edge,'' in \emph{ACM SIGARCH Computer Architecture News}, vol.~45,
  no.~1.\hskip 1em plus 0.5em minus 0.4em\relax ACM, 2017, pp. 615--629.

\bibitem{eshratifar2018energy}
A.~E. Eshratifar and M.~Pedram, ``Energy and performance efficient computation
  offloading for deep neural networks in a mobile cloud computing
  environment,'' in \emph{Proceedings of the 2018 on Great Lakes Symposium on
  VLSI}.\hskip 1em plus 0.5em minus 0.4em\relax ACM, 2018, pp. 111--116.

\bibitem{li2018edge}
E.~Li, Z.~Zhou, and X.~Chen, ``Edge intelligence: On-demand deep learning model
  co-inference with device-edge synergy,'' in \emph{Proceedings of the 2018
  Workshop on Mobile Edge Communications}.\hskip 1em plus 0.5em minus
  0.4em\relax ACM, 2018, pp. 31--36.

\bibitem{satyanarayanan2009case}
M.~Satyanarayanan, V.~Bahl, R.~Caceres, and N.~Davies, ``The case for vm-based
  cloudlets in mobile computing,'' \emph{IEEE pervasive Computing}, 2009.

\bibitem{ha2013just}
K.~Ha, P.~Pillai, W.~Richter, Y.~Abe, and M.~Satyanarayanan, ``Just-in-time
  provisioning for cyber foraging,'' in \emph{Proceeding of the 11th annual
  international conference on Mobile systems, applications, and
  services}.\hskip 1em plus 0.5em minus 0.4em\relax ACM, 2013, pp. 153--166.

\bibitem{ha2014towards}
K.~Ha, Z.~Chen, W.~Hu, W.~Richter, P.~Pillai, and M.~Satyanarayanan, ``Towards
  wearable cognitive assistance,'' in \emph{Proceedings of the 12th annual
  international conference on Mobile systems, applications, and
  services}.\hskip 1em plus 0.5em minus 0.4em\relax ACM, 2014, pp. 68--81.

\bibitem{lim2019federated}
W.~Y.~B. Lim, N.~C. Luong, D.~T. Hoang, Y.~Jiao, Y.-C. Liang, Q.~Yang,
  D.~Niyato, and C.~Miao, ``Federated learning in mobile edge networks: A
  comprehensive survey,'' \emph{arXiv preprint arXiv:1909.11875}, 2019.

\bibitem{teerapittayanon2017distributed}
S.~Teerapittayanon, B.~McDanel, and H.-T. Kung, ``Distributed deep neural
  networks over the cloud, the edge and end devices,'' in \emph{2017 IEEE 37th
  International Conference on Distributed Computing Systems (ICDCS)}.\hskip 1em
  plus 0.5em minus 0.4em\relax IEEE, 2017, pp. 328--339.

\bibitem{huang2019deepar}
Y.~Huang, F.~Wang, F.~Wang, and J.~Liu, ``Deepar: A hybrid device-edge-cloud
  execution framework for mobile deep learning applications,'' 2019.

\bibitem{lin2019cost}
B.~Lin, Y.~Huang, J.~Zhang, J.~Hu, X.~Chen, and J.~Li, ``Cost-driven offloading
  for dnn-based applications over cloud, edge and end devices,'' \emph{arXiv
  preprint arXiv:1907.13306}, 2019.

\end{thebibliography}



\end{document}